\documentclass{aa}
\usepackage{graphicx}
\usepackage{graphicx}
\usepackage{txfonts}
\usepackage{color}

\usepackage{flushend}
\usepackage{epsfig}
\usepackage{epstopdf}
\usepackage{natbib}
\usepackage{pslatex}
\usepackage{multicol}

\def\e\citeppo{$E_{\rm p}$ }

\def\sax{ Beppo SAX }

\begin{document}
\title{Cosmology with  gamma-ray bursts: I. The Hubble diagram through the calibrated $E_{\rm p,i}$ -- $E_{\rm iso}$  correlation}

\titlerunning{ GRB Hubble diagram }
\authorrunning{Demianski et al.}

\author{Marek Demianski$^{1,2}$,  Ester Piedipalumbo$^{3,4}$, Disha Sawant$^{5,6}$, Lorenzo Amati$^{7}$}

\institute{$^{1}$Institute for Theoretical Physics, University of Warsaw, Pasteura 5, 02-093 Warsaw, Poland. \\
$^{2}$Department of Astronomy, Williams College, Williamstown, MA 01267, USA.\\
$^{3}$Dipartimento di Fisica, Universit\`{a} degli Studi di Napoli Federico II, Compl. Univ. Monte S. Angelo, 80126 Naples, Italy\\
\email{ester@na.infn.it}. \\
$^{4}$ I.N.F.N., Sez. di Napoli, Compl. Univ. Monte S. Angelo, Edificio 6, via Cinthia, 80126 - Napoli, Italy.\\
$^{5}$Dipartimento di Fisica e Scienze della Terra, Universit\`{a} degli Studi di Ferrara.\\
$^{6}$Department of Physics, University of Nice Sophia Antipolis, Parc Valrose 06034, Nice Cedex2, France\\
$^{7}$INAF-IASF, Sezione di Bologna, via Gobetti 101, 40129 Bologna, Italy. }

\abstract
{Gamma-ray bursts (GRBs) are the most energetic explosions in the Universe. They are detectable up to very high
redshifts. They may therefore be used to study the expansion rate of the Universe and to investigate the observational properties of
dark energy, provided that empirical correlations between spectral and intensity properties are appropriately calibrated.}
{We used the type Ia supernova (SN) luminosity distances to calibrate the correlation between the peak photon energy,
$E_{\mathrm{p, i}}$, and the isotropic equivalent radiated energy, $ E_{\mathrm{iso}}$ in GRBs. With this correlation, we tested the reliability
of applying these phenomena to measure cosmological parameters and to obtain indications on the basic properties and evolution of dark energy.}
{Using 162 GRBs with measured redshifts and spectra as of the
end of 2013, we applied a local regression technique to calibrate the
$E_{\mathrm{p, i}}$--$E_{\mathrm{iso}}$ correlation against the
type Ia SN data to build a calibrated GRB Hubble diagram. We tested the possible redshift dependence of the
correlation and its effect on the Hubble diagram. Finally, we used the GRB Hubble diagram to investigate the dark energy
EOS. To accomplish this, we focused on the so-called Chevalier-Polarski-Linder (CPL) parametrization of the dark energy EOS and implemented 
the Markov chain Monte Carlo (MCMC) method to efficiently sample the space of  cosmological
parameters.}
 {Our analysis shows once more that the $E_{\mathrm{p, i}}$--$E_{\mathrm{iso}}$ correlation has no significant redshift
dependence. Therefore the high-redshift GRBs can
be used as a cosmological tool to determine the basic cosmological parameters and to test different models of dark energy in the 
redshift region ($z\geqslant 3$), which is unexplored by the SNIa and baryonic acoustic
oscillations data. Our updated calibrated Hubble diagram of GRBs provides some marginal indication (at $1\sigma$ level) of
an evolving dark energy EOS. A significant enlargement of the GRB sample and improvements in the accuracy of the
standardization procedure is needed to confirm or reject, in combination with forthcoming measurements of other
cosmological probes, this  intriguing and potentially very relevant indication.}

 
\date{}
 
\maketitle

\section{Introduction}

\begin{figure}
\centerline
{\includegraphics[width=5.cm,height=4cm]{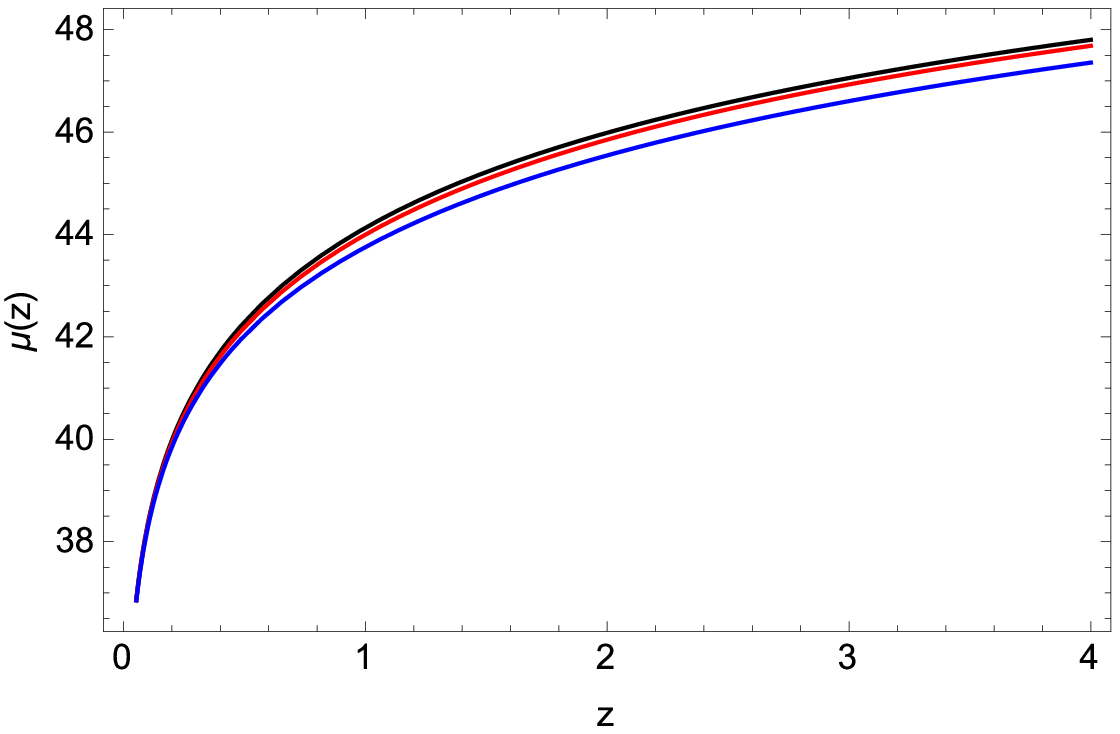}}
%\hspace{0.02\linewidth}
\centerline
{\includegraphics[width=5.1cm,height=4cm]{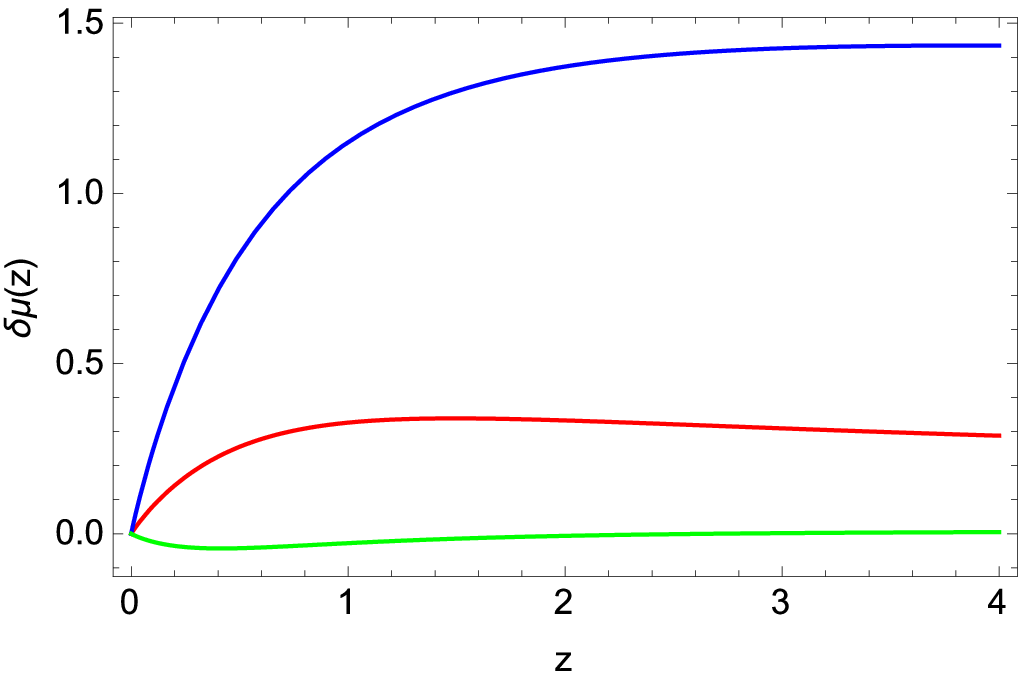}}
\caption{Dependence of the distance modulus on EOS of dark energy. Upper panel: distance modulus $\mu(z)$ for different values of the EOS parameters. 
The black line represents the standard flat  $\Lambda$CDM model with $\Omega_{m}=0.3$, $h=0.7$, $w_0=-1$, $w_1=0$. The other curves correspond to 
different values of the CPL  parameters $w_0$ and $w_1$. Bottom panel: z dependence of the percentage error in the distance modulus between the 
fiducial  $\Lambda$CDM model and different flat  CPL models, as in the left panel.}
\label{evol}
\end{figure}

\begin{figure}
\centerline
{\includegraphics[width=6.cm,height=5cm]{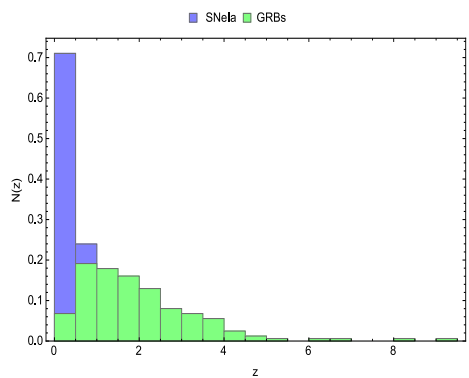}}
\caption{Redshift distribution of our sample of 162 long GRBs/XRFs and the Union 2.1  SNIa dataset.}
\label{nall}
\end{figure}

Starting at the end of the 1990s, observations of high-redshift supernovae of type Ia (SNIa)
revealed that the Universe is now expanding at an accelerated rate  (\citep[see e.g.][]{perl98, per+al99, Riess,Schmidt98, Riess07,SNLS,Union2}).This surprising
result has been independently confirmed by
statistical analyses of observations of small-scale temperature anisotropies of the cosmic microwave background radiation (CMB)
\citep[][]{WMAP3,PlanckXXVI,PlanckXIII}. It is usually assumed that the observed accelerated expansion is
caused by the so called dark energy, a cosmic medium with unusual properties. The pressure of dark energy $p_{de}$ is negative
and is related to the positive energy density of dark energy $\epsilon_{de}$ by $p_{de}=w\,\epsilon_{de}$, where the
proportionality coefficient, that is, the equation of state (EOS), $w$, is negative ($w<- 1/3$). According to current estimates,
about 70\% of the matter energy in the Universe is in the form of dark energy, so that today dark energy is the dominant
component in the Universe. The nature of dark energy is, however, not known. The models of dark energy proposed so far can be
divided into at least three groups: a) a non-zero cosmological constant (see for instance \citep[][]{carroll01}), in this case
$w=-1$, b) a potential energy of some not yet discovered scalar field (see for instance \citep[][]{SF}), or c) effects connected with
the inhomogeneous distribution of matter and averaging procedures (see for instance \citep[][]{clark}). In the last two cases, in
general, $w\not= -1$ and is not constant, but depends on the redshift $z$. To test whether and how $w$ changes with
redshift, it is necessary to use more distant objects. It is commonly argued that since the dark energy density term
becomes sub--dominant (with respect to the dark matter) at $ z \gtrsim 0.5$ \citep[][]{Riess04}, it is not important to study its EOS at earlier epochs. 
However, this argument is unreliable: even within the simplest model, the dark energy contributes nearly $\simeq 10\%$ of the overall 
cosmic energy density at $z \simeq 2$ and
strongly alters the deceleration parameter with
the
cosmological constant. Moreover, for several observables the sensitivity to the dark energy equation of
state increases at high redshifts. In Fig. \ref{evol} we explore this aspect following a simplified approach, considering the
modulus of distance $\mu(z)$ as observable: we fixed a flat $\Lambda$CDM fiducial cosmological model, constructing the
corresponding $\mu_{fid}(z, \rm{ \theta})$, and plot the percentage error on the distance modulus with respect to different
corresponding functions evaluated in the framework of a flat CPL quintessence model \citep[][]{CP01,L03}. We selected
$\Omega_m=0.3$ and $h=0.7$ and varied the dark EOS parameters $w_0$ and $w_1$. It is worth noting that a higher sensitivity is
reached only for $z\gtrsim 2$. Therefore, investigating the cosmic expansion of the
Universe also
beyond these redshifts remains a fundamental probe of dark energy. In this scenario, new possibilities opened up when gamma-ray bursts
were discovered at higher redshifts. The current record is at $z=9.4$ (\citep[][]{Tanvir09, Salvaterra,Cucchiara}). It is worth noting that
the photometric redshift on GRB 090429B is quite high, especially on the low side; GRB 090423, for which a spectroscopic redshift
is available, is
much better determined. GRBs span a redshift
range better suited
for probing dark energy than the SNIa range, as shown
in Fig. (\ref{nall}) , where we compare the distribution in redshift of our sample of 162 long GRBs/XRFs with the Union 2.1 SNIa
dataset.

Gamma-ray bursts are  enigmatic objects, however. First of all, the mechanism that is
responsible for releasing the high amounts of energy that a typical GRB emits is not yet completely known, and only some
aspects of the progenitor models are well established, in particular that long GRBs are produced by core-collapse events, see for instance
\citep[][]{Meszaros06} and \citep[][]{Vedrenne & Atteia}. Despite these difficulties, GRBs are promising objects for studying the
expansion rate of the Universe at high redshifts. One of the most important aspects of the observational property of long GRBs
is that they show several correlations between spectral and intensity
properties (luminosity, radiated energy).
Here we consider the correlation between the observed photon energy of the peak spectral flux, $E_{\mathrm{p, i}}$, which
corresponds to
the peak in the $\nu F_{\nu}$ spectra, and the isotropic equivalent radiated energy $E_{\rm iso}$ (e.g., \citep[][]{Amati02,Amati06}),
\begin{eqnarray}
 \label{eqamati}
  \log \left(\frac{E_{\rm iso}}{1\;\mathrm{erg}}\right)  &=& b+a \log  \left[
    \frac{E_{\mathrm{p,i}} }{300\;\mathrm{keV}}
  \right]\,,
\end{eqnarray}
where $a$ and $b$ are constants, and $E_{\mathrm{p, i}}$ is the spectral peak energy in the GRB cosmological rest--frame,
which can be derived from the observer frame quantity,
$E_{\mathrm{p}}$, by $ E_{\mathrm{p, i}} =E_{\mathrm{p}} (1+z)$.
 This correlation not only provides constraints for the model of the prompt emission, but also naturally suggests that GRBs
can be used as distance indicators.  The isotropic equivalent energy $E_{\rm iso}$ can be calculated from the bolometric fluence $S_{bolo}$ as
\begin{equation}
E_{\rm iso}=4\pi{d^2}_{L}(z, cp)S_{bolo}(1+z)^{-1},
\end{equation}
where $d_{L}$ is the luminosity distance and $cp$ denotes the set of parameters that specify the
background cosmological model. It is clear that  to be able to
use GRBs as distance indicators, it is necessary to consistently calibrate this correlation.
Unfortunately, owing to the lack of GRBs at very low redshifts, the calibration of GRBs is more difficult than that of SNIa, and several calibration
procedures have been proposed so far (see for instance (\citep[][]{CCD,MEC11,MECP12,post14,Dai04, Liu15}).
We here apply a local regression technique to determine the correlation parameters $a$ and $b$, using the recently updated SNIa
sample and  to construct a new calibrated GRB Hubble diagram that can be used for cosmological investigations.
We then use this calibrated GRB Hubble diagram to investigate the cosmological parameters through the Markov chain Monte Carlo technique (MCMC), which 
simultaneously computes the full posterior probability density functions of all the parameters. The structure of the paper is as
follows. In Sect. 2 we describe the methods used to fit the $E_{\rm p,i}$ -- $E_{\rm iso}$ correlation and construct the calibrated GRB Hubble diagram.
In Sect. 3 we present our cosmological constraints and
investigate the possible redshift dependence and Malmquist--like bias effects. In Sect. 4,
as an additional self--consistency check,  we apply the Bayesian method for the non--calibrated $E_{\rm p,i}$ -- $E_{\rm iso}$
correlation and simultaneously extract the correlation coefficients and the cosmological parameters of the
model.  Section 5 is devoted to the discussion of our main results and conclusions.

\section{Standardizing GRBs and constructing the Hubble diagram} 

The GRBs $\nu$ $F_{\nu}$ spectra are well modeled by a 
phenomenological smoothly broken power law, characterized by a low-energy index, $\alpha$, a high-energy index, $\beta$, and a 
break energy $E_0$. They show a peak corresponding to a specific (and observable) value of the photon energy $E_p = E_0 (2+ 
\alpha)$. For GRBs with measured spectrum and redshift it is possible to evaluate the intrinsic peak energy, $E_{\mathrm{p, i}}= 
E_p (1+z)$ and the isotropic equivalent radiated energy \[E_{\rm iso}=4\pi{d^2}_{L}(z, cp)(1+z)^{-1}\int^{10^4/1+z}_{1/1+z} E 
N(E) dE,\] where $N(E)$ is the Band function:

\begin{eqnarray}
\label{eq:recursive}
\resizebox{\linewidth}{!}{$\displaystyle
N(E) = 
\left\{ \begin{array}{ll}
A\left( \frac{E}{100 KeV}\right)^{\alpha} \exp\left(\frac{-E}{E_0}\right)  & \mbox{if }  \left(\alpha-\beta \right) E_0 \geq 0\\
A\left( \frac{\left(\alpha-\beta\right)E}{100 KeV}\right)^{\alpha-\beta} \exp\left(\alpha-\beta\right)\left(\frac{E}{100 KeV}\right)^{\beta} &  \mbox{if }\left(\alpha-\beta\right) E_0\leq E\nonumber\\
\end{array}\right.
$}
\end{eqnarray}

$E_{\rm iso}$ and $E_{\mathrm{p, i}}$ span several orders of magnitude (therefore GRBs cannot be considered standard candles), 
 and show distributions approximated by Gaussians plus a tail at low energies. A strong correlation between these two quantities 
 was initially discovered in a small sample of \sax{} GRBs with known redshifts
(\citep[][]{Amati02}) and confirmed afterward by HETE--2 and SWIFT observations \citep[][]{Lamb05, Amati06}. Several analyses of
the $E_{\mathrm{p, i}}$-$E_{\mathrm{iso}}$ plane of GRBs showed that different classes of GRBs exhibit different behaviors, and while normal long
GRBs and X--ray flashes (XRF, i.e., particularly soft bursts)  follow this correlation, with the exception of the two 
peculiar sub-energetic GRBs 980425 and 031203 \citep[see, e.g.,][for a discussion on possible explanations]{Amati07},
short GRBs do not. These facts may depend on the
different emission mechanisms and/or geometry involved in different classes of GRBs and makes this correlation a useful tool to
distinguish between them (\citep[][]{Amati06,Antonelli09}).
Although it was the first strong correlation discovered for the GRB observables, until recent years, the $E_{\rm
p,i}$ -- $E_{\rm iso}$ correlation
 was never used for cosmology because it exhibits a significant dispersion around the best-fit power law: the residuals 
follow a Gaussian with a value of $\sigma_{\log E_{\rm p,i}} \simeq 0.2$. This type of additional Poissonian scatter is typically 
quantified by performing a maximum likelihood analysis that takes the variance and the errors on dependent and 
independent variables into account. By measuring $E_{\rm p,i}$ in keV and $E_{\rm iso}$ in $10^{52}$ erg, this method gives 
$\sigma_{\log E_{\rm p,i}}=0.19\pm 0.02$ (\citep[][]{Amati08,Amati09}). However, the recent increase in the efficiency of GRB 
discoveries combined with the fact that the $E_{\rm p,i}$ -- $E_{\rm iso}$ correlation requires only two parameters that are directly 
inferred from observations (this minimizes the effects of systematics and increases the number of GRBs that can be used ) 
makes this correlation an interesting tool for cosmology. Despite the very large number of bursts consistent with this 
correlation, its physical origin is still debated. Some authors claimed that it is strongly affected by instrumental selection 
effects (\citep[][]{nakpi05, bandpree05,Butler07, shamo11}).
 However, many other studies found that such instrumental selection biases, even if they may affect the sample, cannot be 
responsible for the existence of the spectral--energy correlations (\citep[][]{Ghirlanda2008}, \citep[][]{Nava12}). 
 Moreover, a recent time--resolved spectral analysis of GRBs that were detected by the BeppoSAX and Fermi
satellite showed that $E_{\rm p,i}$ correlates with the luminosity 
(\citep[e.g.,][]{Gh10,front12}) and radiated energy (\citep[e.g.,][]{Basak13} ) also 
during the temporal evolution of the 
bursts (and the 
correlation between the spectral peak energy and the evolving flux has been pointed out by, \citep[e.g.,][]{Gol83} based on Konus--WIND data). 
The time--resolved $E_{\rm p,i}$ -- luminosity and $E_{\rm p,i}$ -- $E_{\rm iso}$ correlations within individual GRBs and that their average slope is 
consistent with that of correlations 
defined by the time--averaged spectral properties of different bursts strongly supports the physical origin of the $E_{\rm p,i}$ 
-- $E_{\rm iso}$ correlation. It is very difficult to explain them as a selection or
instrumental effect (\citep[e.g.,][]{Dichiara13,Basak13}), and the predominant
emission mechanism in GRB prompt emission produces a correlation between the spectral peak energy
and intensity (which can be characterized either as luminosity, peak luminosity, or radiated energy). 
In addition to its existence and slope, an important property of the $E_{\rm p,i}$ -- $E_{\rm iso}$
correlation is its dispersion. As 
shown by several works that were based on the so-called jet-breaks, 
in the optical afterglow light
curves of some GRBs ( \citep[e.g.,][]{G04,G06}), 50\% of the extra--Poissonian scatter of the correlation is sometimes probably
a result of the
distribution of jet opening angles.
However, these estimates of jet opening
angles are still unconfirmed and model--dependent and can be made only for a small number of GRBs. 
Other factors that may contribute to the dispersion of the $E_{\rm p,i}$ -- $E_{\rm iso}$ and other
$E_{\rm p,i}$ -- intensity correlations include the jet structure, viewing angles,
detectors sensitivity, and energy band (but see Amati et al. 2009),
the diversity of shock micro--physics parameters, and the magnetization within the emitting ejecta. At the current 
observational and theoretical status, it is very difficult to quantify these single contributions, which seem to act randomly in scattering the 
data around the best fit power--law (\citep[e.g.,][]{Ghirlanda2008, Amati08}).
Importantly, it has been shown (\citep[e.g.,][]{Amati09}) that a small fraction (5--10\%) of the scatter
depends on the cosmological model and parameters assumed for computing $E_{\rm iso}$, which makes 
this correlation a potential tool for cosmology.
 
In this section we show how the $E_{\rm p,i}$ -- $E_{\rm iso}$ correlation can be calibrated to standardize  long GRBs and to build a GRB Hubble diagram, which we use 
to investigate different cosmological models at very high redshifts (see also \citep[][]{WangW, Lin15, Lin16,Wangrev}).

\subsection{GRB data}\label{GRBdata}

We used a sample of 162 long GRBs/XRFs as of September 2013 taken from the updated compilation of spectral and
intensity parameters of GRBs by Sawant \& Amati (2016)\nocite{Sawant16}.
The redshift
distribution of this sample covers a broad range, $ 0.03 \leq z\leq  9.3$, which means that it extends far beyond the SNIa range (z $\leq$ 1.7).
These data are of long GRBs/XRFs that are characterized by firm measurements of redshifts and the rest--frame peak energy $E_{\rm
p,i}$. The main contributions come from the joint detections by Swift/BAT and Fermi/GBM or
Konus--WIND, except for the small fraction of events for which Swift/BAT can directly provide $ E_{\rm p,i} $ when it
is in the $\left(15-150\right) $\, keV interval. For events detected by more than one of these detectors, the uncertainties on the
$E_{\rm p,i}$ and $E_{\rm iso}$ take the measurements and uncertainties provided by each individual
detector into account.
In Table~4 we report for each GRB the redshift,
the rest--frame spectral peak energy, E$ _{p,i}$, and the isotropic--equivalent radiated energy, E$ _{iso}$.
As detailed in \citep[][]{Sawant16}, the criterion behind selecting the observations from a particular mission is based on the
conditions summarized below. \\
1. We preferred observations with exposure times of at least two-thirds of the whole event duration.
Hence Konus-WIND and Fermi/GBM were chosen whenever available. For Konus--WIND, the observations were reported
in the official literature (\citep[see e.g. ][]{Ulanov05}), and also in GCN archives when needed. For Fermi/GBM,
the observations were taken from Gruber at al. 2012, from several other papers
(e.g., \citep[][]{G04,G05, FB05}, etc.), and from GCNs.\\
2. The SWIFT BAT observations were chosen when no other preferred mission (Konus--WIND, Fermi/GBM) was able
to provide spectral parameters and the value of E$ _{p,obs}$ was within the energy band of this
instrument. In particular, the E$ _{p,i}$ values derived from BAT spectral analysis were conservatively taken
from the results reported by the BAT team (\citep[][]{Sakamoto08a,Sakamoto08b}). Other BAT E$ _{p,i}$ values reported in the literature
were not considered because they were not confirmed by \citep[][]{Sakamoto08a,Sakamoto08b}, by a refined analysis (\citep[see e.g.][]{cabrera07}),
or because they were based on speculative methods \citep[][]{Butler07}.\\

When available, the Band model \citep[][]{band93} was considered as the cut-off power law, which sometimes overestimates
the value of E$ _{p,i}$.
Finally, the error on any value
was assumed to be not less than 10\%  to account for the instrumental capabilities and calibration uncertainties.

\subsection{Cosmologically independent calibration: local regression of SNIa}

The lack of nearby GRBs  creates  the so-called circularity problem \,:  GRBs can be used as cosmological tools
through the $E_{\rm p,i}$ - $E_{\rm iso}$ correlation, which is based on the cosmological model assumed for the computation
of $E_{\rm iso}$, however.
In principle, this problem could be solved in several ways: it is possible, for instance, to simultaneously  constrain the calibration
parameters $(a, b, \sigma_{int}) \in G $ and the set of cosmological parameters ${\mathbf \theta}\in H$ by considering a likelihood
function defined in the space $G \otimes H$, which allows simultaneously fitting the parameters (e.g., \citep[][]{DOC11}). 
This procedure is implemented in Sect. \ref{full} and compared with the local regression technique. Alternatively, it has been proposed
that the problem might be avoided by considering a sufficiently large number of
GRBs within a small redshift bin centered on any $z$ \citep[][]{G06}.
However, this method, even if quite promising for the  future, is currently unrealistic because of the limited number of
GRBs with measured redshifts. In this section we implement a procedure for calibrating  the $E_{\rm p,i}$ - $E_{\rm iso}$  relation
in a way independent of the
cosmological model by applying  a  local regression technique to estimate the distance modulus $\mu(z)$ from the recently updated
SNIa sample, called Union2.1 (see also \citep[][]{kodama08, Liang08}).
Originally implemented by Cleveland and Devlin (1988), this technique combines the
simplicity of linear regression with the flexibility of nonlinear regression to localized subsets of the data, and reconstructs
a function  describing their behavior in the neighborhood of any point ${\mathbf z_0}$. A low-degree polynomial is
fitted to a subsample  containing a neighborhood of ${\mathbf z_0}$, by using
weighted least-squares with a weight function that quickly decreases with
the distance from ${\mathbf z_0}$. The local regression procedure can be  schematically sketched
as follows:
\begin{enumerate}

\item[1.]{set the redshift $z$ where $\mu(z)$ has to be reconstructed; \\}

\item[2.]{sort the SNIa Union2.1 sample  by increasing value of $|z - z_i|$ and
select the first $n = \alpha {\cal{N}}_{SNIa}$, where  $\alpha$  is a user-selected value and ${\cal{N}}_{SNIa}$ the total number of SNIa; \\}

\item[3.]{apply the weight function\,
\begin{equation}
W(u) = \left \{
\begin{array}{ll}
(1 - |u|^2)^2 & |u| \le 1 \\ ~ & ~ \\ 0 & |u| \ge 1
\end{array}\,,
\right .
\label{eq: wdef}
\end{equation}
where $u = |z - z_i|/\Delta$ and $\Delta$ the highest value of the $|z -
z_i|$ over the previously selected subset;\\ }

\item[4.]{fit a first-order polynomial to the data previously selected and weighted, and use the
zeroth-order term as the best-fit value of the modulus of distance $\mu(z)$, \\ }

\item[5.]{evaluate the error $\sigma_{\mu}$ as the root mean square of the weighted
residuals with respect to the best-fit value. \\ }

\end{enumerate}

 It is worth stressing that both the choice of
the weight function and the order of the fitting polynomial are somewhat arbitrary. Similarly, the value of $n$ to be
used must not be too low to make up a statistically valuable sample, but also not too high to prevent the
use of a low-order polynomial. To check our local regression routine,  we simulated a large catalog
with the same redshift and error distribution as  the Union2.1 survey.
We adopted a quintessence cosmological model with a constant EOS, $w$, and fixed values of the cosmological parameters
$(\Omega_M, w, h)$.  For each redshift value in the Union2.1 sample, we extracted the corresponding modulus of distance from a Gaussian
distribution centered on the theoretically predicted value and with the standard deviation $\sigma = 0.15$. To this value, we added the
value of the error, corresponding to the same relative uncertainty as the data in the Union sample.  This simulated catalog was used
as input to the local regression routine, and  the reconstructed
$\mu(z)$ values were compared to the input ones.

\begin{figure}
\includegraphics[width=8 cm]{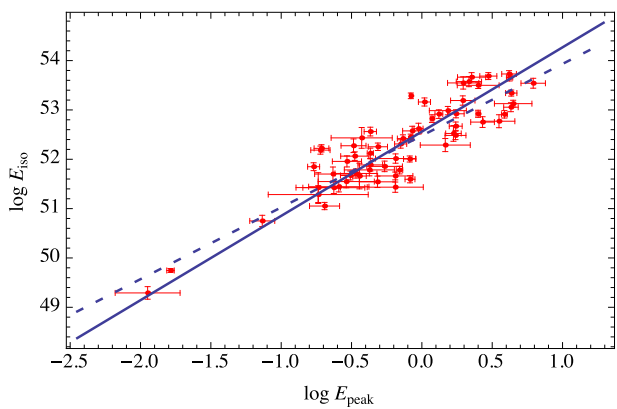}
\caption{Best-fit curves for the  $E_{\rm p,i}$ -- $E_{\rm iso}$ correlation relation superimposed on
the data. The solid and dashed lines refer to the results
obtained with the maximum likelihood ( Reichart likelihood) and weighted $\chi^2$ estimator,
respectively.}
\label{eg-episocorreich}
\end{figure}

Defining the percentage deviation $\epsilon = \frac{
\mu_{w}(z)-\mu_{rec.}(z)}{\mu_{w}(z)}$ with $\mu_{rec.}$ and $\mu_{w}$ the local
regression estimate and the input values, respectively, and averaging
over 500 simulations, we found that the choice
$\alpha = 0.02$ gives $(\delta \mu/\mu)_{rms} \simeq 0.3\%$ with $|\epsilon | \le 1\%$ independently on the redshift $z$. This result
implies that the local regression method
allows correctly reconstructing the underlying distance modulus regardless of
redshift $z$$<$1.4 from the Union SNIa sample. We also tested this results in different cosmological backgrounds by  adopting
an evolving EOS and averaging over five realizations of the mock catalog.
With this efficient way of estimating $\mu(z)$ at
redshift $z$ in a model-independent way, we can now fit the $E_{\rm p,i}$ -- $E_{\rm iso}$ correlation relation, using
the reconstructed $\mu(z)$. We considered only GRBs with $z \le 1.414$  to cover the same redshift range as is spanned by the SNIa data.
To standardize the $E_{\rm p,i}$ -- $E_{\rm iso}$ relation as expressed
by the Eq. (\ref{eqamati}), we need to
fit a data array $\{x_i, y_i\}$ with uncertainties $\{\sigma_{x,i}, \sigma_{y,i}\}$, to a straight line
\begin{equation}
\label{linear} y =b + a x\,,
\end{equation}
and determine the  parameters $(a, b)$. We expect a certain amount of intrinsic additional Poissonian scatter, $\sigma_{int}$, around the best-fit line
that has to be taken into account and determined together with $(a, b)$ by the fitting procedure. We used  a
likelihood,  implemented by  Reichart  \citep[][]{Reichart01}, that
solved this problem of fitting data
 that are affected by  extrinsic scatter in addition  to the intrinsic uncertainties along both axes:
\begin{eqnarray}
L_{Reichart}(a, b, \sigma_{int}) & = & \frac{1}{2} \frac{\sum{\log{(\sigma_{int}^2 + \sigma_{y_i}^2 + a^2
\sigma_{x_i}^2)}}}{\log{(1+a^2)}} \\ \nonumber &&+ \frac{1}{2} \sum{\frac{(y_i - a x_i - b)^2}{\sigma_{int}^2 + \sigma_{x_i}^2 + a^2
\sigma_{x_i}^2}}\,,\label{eq: deflike}
\end{eqnarray}
where the sum is over the ${\cal{N}}$ objects in the sample. We note that this maximization was performed in the
two-parameter space $(a, \sigma_{int})$ since $b$ may be calculated analytically by solving the equation  $\displaystyle
{\frac{\partial }{\partial b}L(a, b, \sigma_{int})=0}$, we obtained
\begin{equation}
b = \left [ \sum{\frac{y_i - a x_i}{\sigma_{int}^2 + \sigma_{y_i}^2
+ a^2 \sigma_{x_i}^2}} \right ] \left [\sum{\frac{1}{\sigma_{int}^2 + \sigma_{{y_i}}^2 + a^2 \sigma_{x_i}^2}} \right ]^{-1}\,. \label{eq:calca}
\end{equation}
To quantitatively estimate the goodness of this fit, we used
the median and root mean square of the best-fit residuals, defined
as $\delta = y_{obs} - y_{fit}$.
To quantify the uncertainties of some fit parameter $p_i$, we
evaluated the marginalized likelihood ${\cal{L}}_i(p_i)$  by
integrating over the other parameters. The median value for the
parameter $p_i$ was then found by solving
\begin{equation}
\int_{p_{i,min}}^{p_{i,med}}{{\cal{L}}_i(p_i) dp_i} = \frac{1}{2}
\int_{p_{i,min}}^{p_{i,max}}{{\cal{L}}_i(p_i) dp_i} \ . \label{eq:
defmaxlike}
\end{equation}
The $68\%$ ($95\%$) confidence range $(p_{i,l}, p_{i,h})$ was then
found by solving \citep[e.g., ][]{dagostini}
\begin{equation}
\int_{p_{i,l}}^{p_{i,med}}{{\cal{L}}_i(p_i) dp_i} = \frac{1 -
\varepsilon}{2} \int_{p_{i,min}}^{p_{i,max}}{{\cal{L}}_i(p_i) dp_i}
\ , \label{eq: defpil}
\end{equation}
with $\varepsilon = 0.68$ and $\varepsilon = 0.95$ for the $68\%$
and $95\%$ confidence level.
 The maximum likelihood values of $a$ and $\sigma_{int}$ are $a=1.75^{+0.18}_{-0.16}$ and $\sigma_{int}=0.37^{+0.07}_ {-0.08}$.
In Fig. \ref{eg-episocorreich} we show the correlation  $\log{E_{\rm p,i}}$ -- $\log{E_{\rm iso}}$ . The solid line is the best
fit obtained using the Reichart likelihood, and the dashed line is the best fit obtained by the weighted $\chi^2$ method.

\begin{figure}
\includegraphics[width=8 cm, height=4 cm]{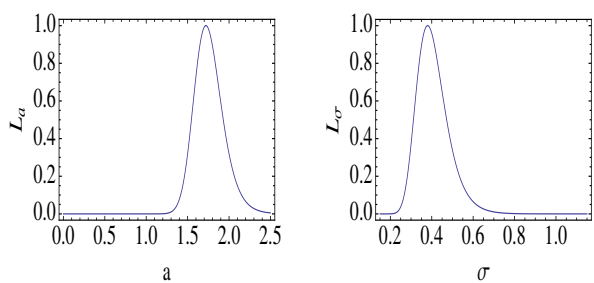}
\caption{Marginalized likelihood functions:  the likelihood function $ {\cal{L}}_{a}$ is obtained by marginalizing  over
$\sigma_{int}$; and the likelihood function,  ${\cal{L}}_{\sigma_{int}}$, is obtained by marginalizing over $a$.}
\label{likelihood_reich}
\end{figure}

The marginalized likelihood functions are shown in Fig. \ref{likelihood_reich}.

\begin{figure}
\includegraphics[width=8.5 cm]{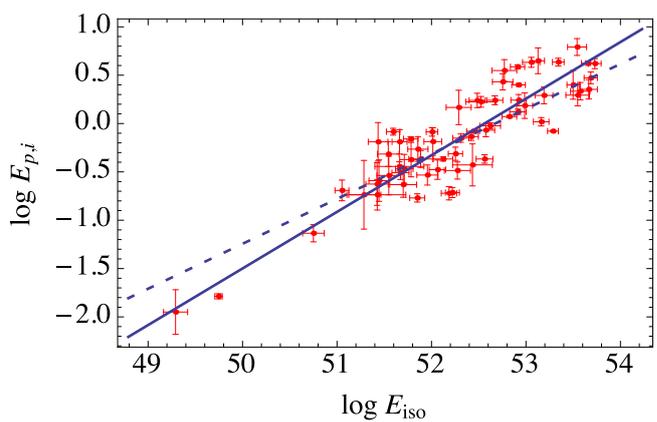}
\caption{Best-fit curves for the  $E_{\rm iso}$--$E_{\rm p,i}$  correlation relation superimposed on
the data. The solid and dashed lines refer to the results
obtained with the maximum likelihood (Reichart likelihood) and weighted $\chi^2$ estimator,
respectively.}
\label{amatinverse}
\end{figure}

As noted above, $b$ can be evaluated analytically. We obtained $b=52.53\pm 0.02$. It is worth noting that in the literature results for the 
inverse correlation are commonly reported, that is, the correlation $E_{\rm iso}$--$E_{\rm p,i}$: using the local regression technique and 
the Reichart likelihood, we also obtained this inverse calibration. We obtained $a^{inverse}=0.58^{+0.07}_{-0.05}$, $\sigma^{inverse}_{int}=0.24^{+0.04}_{0.03}$.  
In Fig. \ref{amatinverse} we plot the best-fit curves for the $E_{\rm iso}$--$E_{\rm p,i}$  correlation relation superimposed on the data.

\begin{figure}
\hspace{-0.0002em}
\centerline
{\includegraphics[width=7.8 cm]{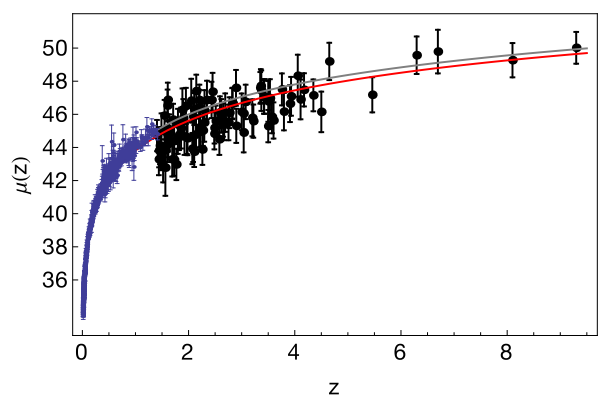}}
\caption{ Calibrated GRB Hubble diagram (black filled diamond) up to very high values of redshift, as constructed
by applying the local regression technique with the Reichart likelihood: we show over--plotted  the behavior of the theoretical
distance modulus $\mu(z)= 25 + 5 \log{d_L(z)}$ corresponding to  the favored  fitted CPL model (red line), with $w_0= -0.29$,
$w_1=-0.12$, $h=0.74$, $\Omega_m=0.24$, and the standard $\Lambda$CDM model
(gray line) with $\Omega_m=0.33$, and $h=0.74$.They are defined by Eqs. (\ref{eq:defmu}) and (\ref{eqlum}).
The blue filled points at lower redshift are the SNIa data.}
\label{hdgrbamati}
\end{figure}

\subsubsection{Constructing the Hubble diagram}

After fitting the  correlation  and estimating its parameters, we used them
to construct the  GRB Hubble diagram. We recall that the luminosity distance
of a GRB with the redshift $z$ may be computed as
\begin{equation}\label{lumdist}
d_L(z) = \left( \frac{E_{\rm iso}(1 + z)}{4 \pi  S_{bolo}}\right)^{1/2}.
\end{equation}
The uncertainty of $d_L(z)$ was then estimated through the propagation of the measurement errors on the involved
quantities. In particular, recalling that our correlation relation  can be written as a linear relation, as in  Eq.
(\ref{linear}), where $y$$=$$E_{iso}$ is  the distance dependent quantity, while $x$ is not, the error on the distance dependent
quantity $y$ was estimated as
\begin{equation}
\sigma(y) = \sqrt{a^2 \sigma^2_{x}+\sigma_{a}^2 x^2 +\sigma_{b}^2+ \sigma_{int}^2},
\label{eq:siglogy}
\end{equation}
where $\sigma_b$ is properly evaluated through the Eq. (\ref{eq:calca}), which implicitly  defines b as a function of $a$ and $\sigma_{int}$,
and is then added in quadrature to the uncertainties of the other terms entering
Eq.(\ref{lumdist}) to obtain the total uncertainty. The distance modulus $\mu(z)$ is
easily obtained from its standard definition\,\begin{equation}
\mu(z) = 25 + 5 \log{d_L(z)}\,, \label{eq:defmu}
\end{equation}
with its uncertainty obtained again by error propagation:
\begin{equation}
\sigma^2_{\mu} = \left(\frac{5}{2}\sigma^2(y)\right)^2+ \left(\frac{5}{2 \ln 10 }\frac{\sigma_{S_{bolo}}}{S_{bolo}}\right)^2\label{sigmamu}\,.
\end{equation}

\begin{figure}
\includegraphics[width=7 cm]{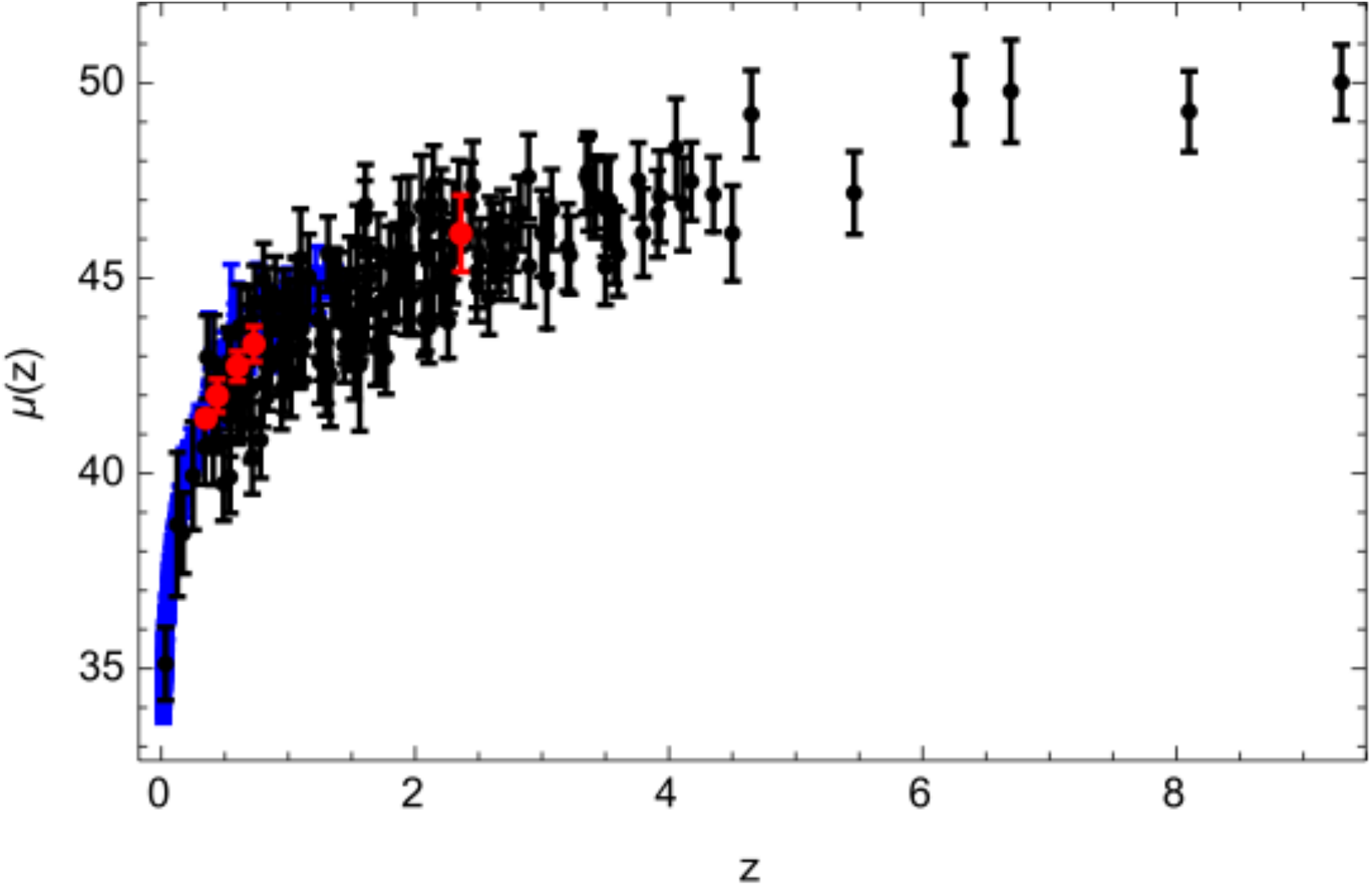}
\caption{GRB Hubble diagram (black points) is compared with the SNIa Hubble diagram (blue points) and with BAO data (red points). }
\label{all}
\end{figure}

 We finally estimated the distance modulus for each $i$\,-\,th
GRB in the sample at redshift $z_i$ to build the Hubble diagram plotted in Fig. \ref{hdgrbamati}. We
refer to this data set as the  calibrated GRB Hubble diagram below since to compute the distances, the Hubble diagram
 we relied  on the calibration was based on the SNIa Hubble diagram. The derived distance moduli are divided into two
subsets,  listed in Tables \ref{calibratedHdlow} and \ref{calibratedHdhigh}. Table 4 lists GRBs with $z\leq 1.46$, the
same redshift range as for known SNIa, and Table 5 lists GRBs with $z \geq 1.47$. In Fig. (\ref{all}) we finally
compare the GRB Hubble diagram (black points) with the SNIa Hubble diagram (blue points) and with BAO data (red points).

\section{Cosmological constraints derived from  the calibrated GRB Hubble diagram }

Here we illustrate  the possibilities of using the GRB Hubble diagram to constrain the cosmological  models and investigate the dark energy EOS.
Within the standard homogeneous and isotropic cosmology, the dark energy appears in the Friedman equations:
\begin{equation}
\frac{\ddot{a}}{a} = - \frac{4 \pi G}{3} \ \left(\rho_M + \rho_{de} + 3 p_{de}\right) \ , \label{eq: fried1}
\end{equation}
\begin{equation}
H^2 +\frac{k}{a^2}= \frac{8 \pi G}{3} (\rho_M + \rho_{de}) \,, \label{eq: fried2}
\end{equation}
where $a$ is the scale factor, $H = \dot{a}/a$ the Hubble parameter, $\rho_M$ the density of matter,
$\rho_{de}$ the density of dark energy, $p_{de}$ its pressure, and the dot denotes the time derivative.
The continuity equation for any component of the cosmological fluid is \,
\begin{equation}
\frac{\dot{\rho_i}}{\rho_i} = - 3 H \left(1 + \frac{p_i}{\rho_i}\right) = - 3 H  \left[1 + w_i(t)\right]\,,
\label{eq: continuity}
\end{equation}
where the energy density is $\rho_i$, the pressure $p_i$, and the EOS of the $i-th$ component is defined by
$\displaystyle{w_i=\frac{p_i}{\rho_i}}$. The standard non-relativistic matter has $w=0$, and the cosmological constant has
$w=-1$. The dark energy EOS and other constituents of the Universe determine the Hubble function $H(z)$
and any derivations of it that are needed to obtain the observable quantities. When only
matter and dark energy are present, the Hubble function is given by
\begin{eqnarray}\label{heos}
 && H(z,{\rm\theta}) = H_0 \sqrt{(1-\Omega_m) g(z, {\rm \theta})+\Omega_m (z+1)^3+ \Omega _k (z+1)^2}\,,\nonumber\\
\end{eqnarray}
where $\displaystyle{g(z)=\exp (3 \int_0^z \frac{w(x,{\rm \theta})+1}{x+1} \, dx)}$,  and $w(z,{\rm \theta})$ describes
the  dark energy EOS, characterized by $n$ parameters ${\rm \theta}=(\theta_1, \theta_2..,\theta_n)$.  We limit our analysis below to the CPL  parametrization,
 \begin{equation}
w(z) =w_0 + w_{1} z (1 + z)^{-1} \,,
\label{cpleos}
\end{equation}
where $w_0$ and $w_1$ are constant parameters and they
represent the $w(z)$ present value and its
overall time evolution, respectively \citep[][]{CP01,L03}.
  If we introduce  the dimensionless Hubble parameter
$E(z,{\rm\theta}) $ :
\begin{equation}
\begin{small}
E(z,{\rm\theta})=\sqrt{(z+1)^2 \Omega _k + (z+1)^3 \Omega _m+\Omega _{\Lambda } e^{-\frac{3 w_1 z}{z+1}} (z+1)^{3 \left(w_0+w_1+1\right)}}\,,
 \label{eq:ezfull}
 \end{small}
\end{equation}
 we can define the luminosity distance and the modulus of distance. Actually the luminosity distance id given by:
\begin{equation}
d_L(z,{\rm\theta}) = d_H (1+z)d_{M}(z,{\rm\theta})\,, 
\end{equation}
where $d_H= \frac{c}{H_0}$, $d_{M}(z,{\rm\theta})$ is the transverse co-moving distance  and it is defined as
\begin{equation}
%\begin{small}
 d_{M}(z,{\rm\theta}) =\left \{
\begin{array}{ll}
\frac{d_H}{\sqrt{\Omega_k}} \sinh \frac{d_C(z,{\rm\theta})}{d_H}& \Omega_k > 0, \\ ~ & ~ \\   \frac{d_H}{\sqrt{|\Omega_k|}} \sin \frac{d_C(z,{\rm\theta})}{d_H}&  \Omega_k < 0,\\ ~ & ~ \\   d_C(z,{\rm\theta})& \Omega_k = 0,\\
\end{array}
\right.\,
\label{eqlum}
%\end{centering}
%\end{small}
\end{equation}
%\end{widetext}
being $d_C(z,{\rm\theta})$  the co-moving distance:
\begin{equation}
%\begin{small}
d_C(z,{\rm\theta})=d_H \int_0^z \frac{d\zeta}{E(\zeta,{\rm\theta})}\,.
 % \end{small}
\end{equation}
Therefore we can define the modulus of distance $\mu_{th}(z, {\rm\theta})$:
\begin{equation}
\mu_{th}(z, {\rm\theta})= 25 + 5 \log{  d_L(z,{\rm\theta}) }\,.
\end{equation}
The standard $\Lambda$CDM model corresponds to $w_0=-1$, $w_1=0$.

%It is worth noting that the  calibrated GRB Hubble diagram spans an \textcolor[rgb]{1,0.501961,0}{{\it optimal\LEt{see abstract}}}  redshift range, as far as the sensitivity of %the observablequantities on the cosmological parameters is concerned, in particular to the dark energy EOS.
%\subsection{Statistical analysis}
To constrain the parameters specifying different cosmological models, we  maximized the likelihood function
${\cal{L}}({\rm\theta}) \propto \exp{[-\chi^2({\rm\theta})/2]}$, where ${\rm\theta}$ indicates the set of cosmological parameters
and the  $\chi^2({  \rm\theta})$ was defined as usual by 
\begin{eqnarray}
\chi^2({\rm\theta}) & = & \sum_{i = 1}^{{\cal{N}}_{GRBs}}{\left [ \frac{\mu_{obs}(z_i) - \mu_{th}(z_i, {\rm\theta})}{\sigma_i} \right ]^2} \ .
\label{defchiGRB}
\end{eqnarray}
Here, $\mu_{obs}$ and $\mu_{th}$ are the observed and theoretically predicted values of the distance modulus. The parameter space is
efficiently sampled by using the MCMC method, thus  running five
parallel  chains and using the Gelman-Rubin test to check the convergence  \citep[][]{GR92}. The  confidence levels are estimated from
the likelihood quantiles.
We recall that we performed the analysis assuming a non-zero spatial curvature (not flat $\Lambda$CDM), and only in this case did we take $w=-1$.
To alleviate the strong degeneracy of the curvature parameters and any EOS parameters, we added a Gaussian prior
on $\Omega_k$, centered on the value
provided by the Planck collaboration \citep[][]{PlanckXXVI}, $100\Omega_k^{Planck}=-4.2^{+4.3}_{-4.8}$, and with a dispersion  ten times  of $\sigma_{\Omega_k}^{Planck}$, where $100\sigma_{\Omega_k}^{Planck}=4.5$.
We investigated the CPL parameters assuming a flat universe.
In a forthcoming paper we will present a full cosmological analysis using the high-redshift GRB Hubble diagram to test
different cosmological models, where several  parameterizations of the dark energy EOS will be used and  also different
dark energy scenarios, for instance the scalar field quintessence.  In Table
\ref{tab:fullgrbtab2} we summarize the results of our analysis. There are indications that the dark energy EOS is evolving.

\begin{figure}
\centerline
{\includegraphics[width=7 cm, height= 7 cm]{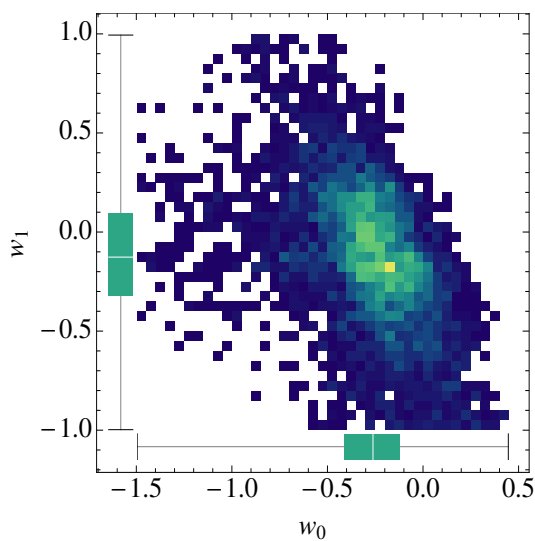}}
\caption{Regions of confidence for the marginalized likelihood function $ {\cal{L}}(w_0,w_1)$. On the axes we also plot the
box-and-whisker diagrams for the $w_0$ and $w_1$ parameters:  the bottom and top of the diagrams are,  as commonly done in the  box-plots,  the 25th
and 75th percentile (the lower and upper quartiles, respectively), and the band near the middle of the box is  the 50th percentile
(the median). It is evident that the $\Lambda$CDM model, corresponding to $w_0=-1$ and $w_1=0$ is not favored.}
\label{figGRBcosparcali}
\end{figure}

\begin{table*}
\caption{Constraints on the cosmological parameters. Columns report the mean $\langle x \rangle$ and median
$\tilde{x}$ values and the $68\%$ and $95\%$ confidence limits. The investigation of the CPL
parametrization for the dark energy  was performed assuming a flat universe (upper side). The analysis
performed by assuming a non-zero spatial curvature is limited to the case  $w=-1$  (non-flat  $\Lambda$CDM). 
The GRB Hubble diagram alone provides $\Omega_k^{median}=  -0.00046,$ in agreement with the CMBR
results.} \label{tab:fullgrbtab2}
\begin{center}
%\scriptsize
\resizebox{10 cm}{!}{
\begin{tabular}{cccccccccc}
\hline
~ & \multicolumn{4}{c}{ Calibrated GRB Hubble diagram and cosmology} \\
\hline
$Id$ & $\langle x \rangle$ & $\tilde{x}$ & $68\% \ {\rm CL}$  & $95\% \ {\rm CL}$ \\
\hline \hline
~ & ~ & ~ & ~ & ~   \\
$\Omega_m$ & 0.24 &0.19 & (0.12, 0.37) & (0.10, 0.58)   \\
~ & ~ & ~ & ~ & ~   \\
$w_0$ & -0.29& -0.26 & (-0.5,-0.1) & (-1.01,0.1)  \\
~ & ~ & ~ & ~ & ~\\
$w_{1}$ &-0.12& -0.13 & (-0.43, 0.19) & (-0.88,0.6)  \\
~ & ~ & ~ & ~ & ~\\
$h$ &0.74& 0.74& (0.69, 0.78) & (0.65,0.8)  \\
~ & ~ & ~ & ~ & ~ \\

\hline \hline
~ & ~ & ~ & ~ & ~   \\
$\Omega_m$ & 0.33 &0.32 & (0.19, 0.49) & (0.12, 0.59)   \\
~ & ~ & ~ & ~ & ~  \\
$\Omega_{\Lambda}$ & 0.66& 0.677 & (0.51, 0.8) & (0.42,0.87)  \\
~ & ~ & ~ & ~ & ~  \\
$h$ &0.74& 0.74& (0.70, 0.77) & (0.66,0.79)  \\
~ & ~ & ~ & ~ & ~ \\

$\Omega_{k}$ &-0.00049& -0.00046 & (-0.007, 0.0064) & (-0.014,0.013)  \\
~ & ~ & ~ & ~ & ~   \\
\hline
\end{tabular}}
\end{center}
\end{table*}

The joint probability for different sets of parameters that characterize the CPL EOS is shown in Fig.
\ref{figGRBcosparcali}.

\section{$E_{\rm iso}$  and $E_{\rm p,i}$ correlation}

\begin{figure}
\centerline
{\includegraphics[width=6.cm,height=4cm]{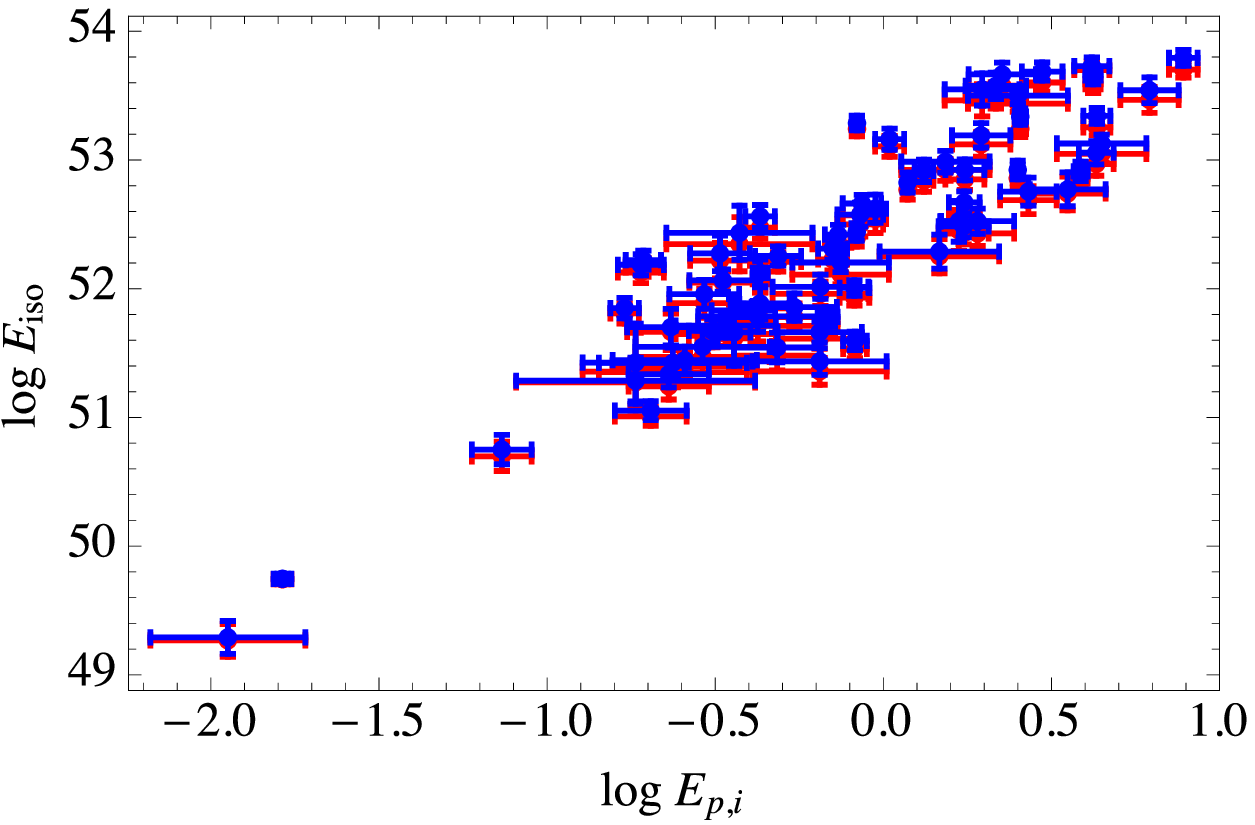}}
\vspace{0.02\linewidth}
%\centerline
\hspace{3em}
{\includegraphics[width=6.2cm,height=4.2 cm]{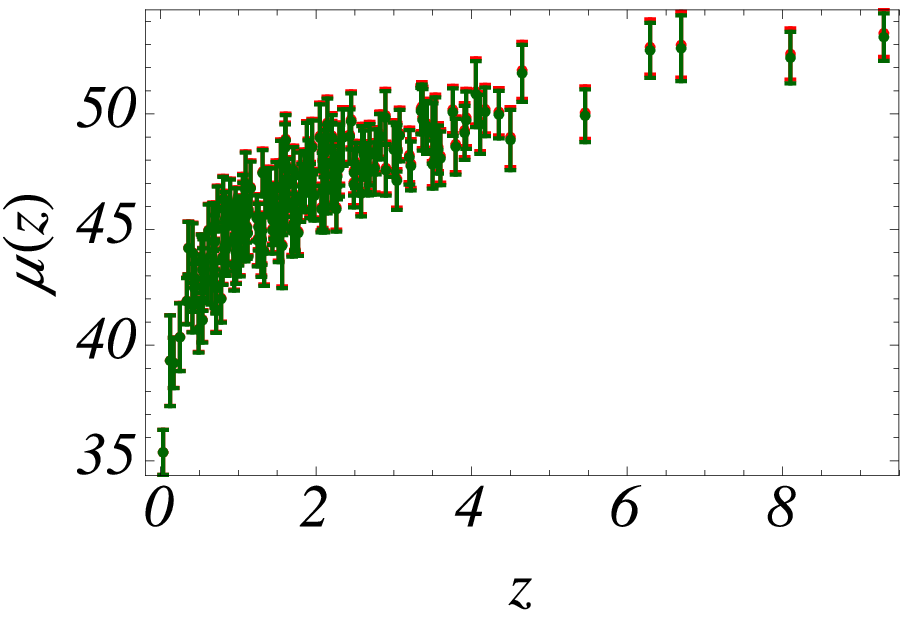}}
\caption{Evolutionary effects on the $E_{\rm iso}$ -- $E_{\rm p,i}$ correlation and the Hubble flow. Upper panel: de-evolved (red points) and  
evolved/original correlation (green points); there is no discernible evolution. Bottom panel:   de-evolved (red points) and   evolved/original 
(green points) Hubble diagram; evolutionary effects of the correlation do not  affect the GRB Hubble diagram.}
\label{Correlevolution}
\end{figure}

Since we here used the calibrated GRB Hubble diagram  to perform the  cosmological investigation described above, we
discuss some arguments about the reliability of using the $E_{\rm iso}-E_{\rm p,i}$  relation for cosmological tasks. For instance,
the calibration method we implemented so far relies  on the underlying assumption that the calibration parameters do not evolve
with redshift. It is worth noting that the problem of the redshift dependence of the GRBs correlations is widely debated in the
literature,
with different conclusions. This problem is intimately connected to the problem of the influence of possible selection effects
or biases on
the observed correlations, see, for instance, \citep[][]{Li08, caballerapetro,slope, Liu15, WangW}. Any answer to these fundamental
questions is far from being settled until more data with known redshifts are available.  In this section, however,  we revisit this
question from an observational point of view: we test  the validity of this commonly adopted working hypothesis and search for any  evidence
of such a redshift dependence. We also investigate possible effects on the GRB Hubble diagram. As a first  simplified approach we considered
two subsamples with a comparable number of bursts, divided according to redshift: a  lower redshift sample  of $97$ bursts
with $z\leq 2$, and a  higher redshift sample of 67 burst with $z>2$.  We estimated the  cosmological parameters  for flat $\Lambda CDM$
and  CPL dark energy EOS cosmological models, considering the two subsamples separately, and compared the results.  Even
when the bursts belonging to these two samples experience different  environment  conditions, we did not find significant indication that
a spurious z-evolution of the slope affects the cosmological fit:  the values of the cosmological parameters derived from these
two samples  are statistically  consistent  within $ 1 \sigma$, as shown in Table \ref{tabcosmofit}. This result indirectly shows
that redshift evolution dependence, even if it exists,  does not undermine the reliability of the GRBs as probes of the cosmological
expansion. Moreover, to investigate this redshift dependence in more detail, that is, how it  affects  $E_{\rm iso}$ and/or $E_{\rm p,i}$
(which we generically denote by $y$), we used two different approaches. First, we evaluated the Spearman rank correlation coefficient,
$C(z,y)$, taking into account that since our sample is not too large, a few points could dominate the final value of the rank correlation,
which would introduce a bias. We applied  a {\it jacknife} re-sampling method by evaluating $C(z, y)$ for $N-1$ samples obtained by excluding
one GRB at a time, and we adopted the mean value and the standard deviation  to estimate $C(z,y)$.  
We found $ C(z,E_{p,i})= 0.299 \pm 0.004$, and $ C(z, E_{iso})= 0.278 \pm 0.004$, which indicates a moderate evolution of the correlation.
The Spearman rank correlation coefficient, however, does not include the errors of  $E_{\rm iso}$ and  $E_{\rm p,i}$, so that we
also implemented an alternative, and completely different,  approach to determine whether and how strongly these variables
are correlated with redshift. We assumed that  the evolutionary functions can be parametrized by a simple power-law
functions: $g_{iso}(z)=\left(1+z\right)^{k_{iso}}$ and $g_{p}(z)=\left(1+z\right)^{k_{p}}$  (see for instance\citep[][]{caballerapetro}),
so that $E_{\rm iso}^{'} =\displaystyle\frac{E_{\rm iso}}{g_{iso}(z)}$ and $E_{\rm p,i}^{'} =\displaystyle\frac{E_{\rm p,i}}{g_{p}(z)}$
are the de-evolved  quantities. In this case,  the effective $E_{\rm iso}-E_{\rm p,i}$ correlation can be written as a 3D correlation:
\begin{eqnarray}
 \label{eqamatievol}
&&\log \left[\frac{E_{\rm iso}}{1\;\mathrm{erg}}\right] = b+a \log  \left[
    \frac{E_{\mathrm{p,i}} }{300\;\mathrm{keV}} \right]+ \left(k_{iso} - a \,k_{p}\right)\log\left(1+z\right)\,.\nonumber\\
\end{eqnarray}
Calibrating this 3D  relation means determining the coefficients ($a$, $ b$, $k_{iso}$, and $k_{p}$) plus the intrinsic scatter $\sigma_{int}$.  
Low values for $k_{iso}$ and $k_{p}$ would indicate a lack of evolution, or at least negligible evolutionary effects. To fit these coefficients,
we constructed a 3D Reichart likelihood, but we consider no error on the redshift of each GRB:
\begin{eqnarray}
 \label{reich3dl}
&&L^{3D}_{Reichart}(a, k_{iso}, k_{p}, b,  \sigma_{int}) =  \frac{1}{2} \frac{\sum{\log{(\sigma_{int}^2 + \sigma_{y_i}^2 + a^2
\sigma_{x_i}^2)}}}{\log{(1+a^2)}}\nonumber \\ &+& \frac{1}{2} \sum{\frac{(y_i - a x_i -(k_{iso}-\alpha) z_i-b)^2}{\sigma_{int}^2 + \sigma_{x_i}^2 + a^2
\sigma_{x_i}^2}}\,,
\end{eqnarray}

where $\alpha= a\, k_{p}$.
As in the $2D$ case, we maximized our likelihood in the space ($a$,  $k_{iso}$, and $\alpha$) since $b$ was evaluated analytically by solving the equation
$\displaystyle
{\frac{\partial }{\partial b}L^{3D}_{Reichart}(a, k_{iso}, k_{p}, b, \sigma_{int})=0}$, we obtain
\begin{equation}
b = \left [ \sum{\frac{y_i - a x_i-(k_{iso}-\alpha) z_i}{\sigma_{int}^2 + \sigma_{y_i}^2
+ a^2 \sigma_{x_i}^2}} \right ] \left [\sum{\frac{1}{\sigma_{int}^2 + \sigma_{{y_i}}^2 + a^2 \sigma_{x_i}^2}} \right ]^{-1}\,. \label{eq:calca}
\end{equation}
We also used the MCMC method and ran five
parallel chains and the Gelman-Rubin convergence test, as  previously explained.
We finally studied the median and $68\% $ confidence range of $k_{iso}$\, and $\alpha$ to test whether the correlation evolves
and noted
that a null value for these parameters is strong evidence for a lack of any evolution. We found that
$a=1.86^{+0.07}_{-0.09}$, $k_{iso}=-0.04\pm 0.1$; $\alpha=-0.02\pm 0.2$\,;
$\sigma_{int}=0.35\pm 0.03$, so that $b= 52.40_{-0.06}^{+0.03}$.  We can safely conclude that the $E_{\rm iso}$ and
$E_{\rm p,i}$ correlation shows, at this stage, weak indications of evolution. In Fig.\ref{Correlevolution} we
plot both the de-evolved and   evolved/original correlation, and  the de-evolved and  evolved/original
Hubble diagram: we do not see any signs of evolution.

\section{Fully Bayesian analysis }\label{full}

In this section we simultaneously  constrain the calibration parameters $(a, b, \sigma_{int})$ and the set of cosmological
parameters by considering the same likelihood function as in Eq. (\ref{eq: deflike}).
Our task is to determine  the multidimensional probability distribution function (PDF) of the
parameters $\{ a,b,\sigma_{\rm int} , {\mathbf p} \} $,
where ${\mathbf p}$ is the ${\cal{N}}$-dimensional vector of
the cosmological parameters. The Amati correlation can be written in the form
\begin{equation}
\log_{10} S_{\rm bol} = a + b \log_{10} E_{\rm p,i} - \log_{10}[4\pi d_L^2(z,{\mathbf p})].
\end{equation}

\begin{figure}
\centerline
{\includegraphics[width=6 cm, height=6 cm]{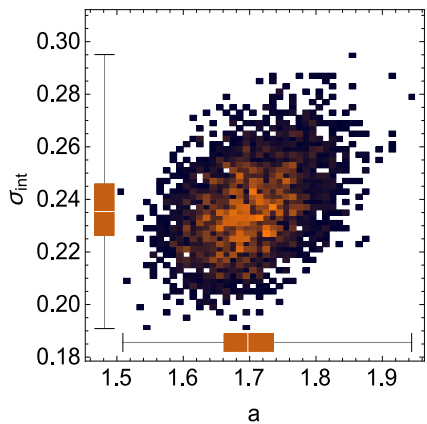}}
\caption{Regions of confidence for the marginalized likelihood function $ {\cal{L}}(a,\sigma)$, obtained by marginalizing over $b$
and the cosmological  parameters. The gray regions indicate the $1\sigma$ (full zone) and $2\sigma$ (empty zone) regions of
confidence. On the axes we also plot the box-and-whisker diagrams for the $a$ and $\sigma_{int}$ parameters:
the bottom and top of the diagrams are  the 25th and 75th percentile (the lower and upper quartiles, respectively), and the band near
the middle of the box is  the 50th percentile (the median).}
\label{figGRBpar}
\end{figure}

We note that  the best-fit zero point $b$ can be analytically expressed as a function of $(a, \sigma_{int})$ when the cosmological parameter
${\bf p}_c$ is specified. A more computationally efficient strategy is to
let $b$  free and add it to the list of quantities to be determined, which thus sums up to ${\cal{N}} + 3$. To efficiently sample the
${\cal{N}} + 3$ dimensional parameter space, we used the MCMC method and ran five
parallel chains and used the Gelman-Rubin convergence test, as described in the previous section. Since we are mainly
interested in the calibration problem and not in constraining the cosmological parameters $(\Omega_M, \Omega_{\Lambda},
h)$, we considered here only  the particular case of a flat $\Lambda$CDM model, as already done in the previous
analysis. The analysis determines, at the same time, the cosmological parameters and the correlation coefficients,
which are listed in Table \ref{tab: fullgrbtab}. Although the calibration procedure is different (since we now
fit for the cosmological parameters as well), it is nevertheless worth comparing this determination of $(a, b,
\sigma_{int})$ with the one  obtained in the previous analysis based on the SNIa sample. It is evident that although
the median values change,  the  $95\%$ confidence levels  are in full agreement so that we cannot find any
statistically significant difference. Figure \ref{figGRBpar} shows the marginalized probability distribution function
of the correlation coefficients of the relation $E_{\rm iso} -E_{\rm p,i}$.

\begin{table*}
\begin{center}
%\scriptsize
\resizebox{17cm}{!}{
\begin{tabular}{cccccccccc}
\hline
~ & \multicolumn{4}{c}{Lower redshift sample} & \multicolumn{4}{c}{Higher redshift sample}\\
\hline
$Id$ & $\langle x \rangle$ & $\tilde{x}$ & $68\% \ {\rm CL}$  & $95\% \ {\rm CL}$ & $\langle x \rangle$ & $\tilde{x}$ & $68\% \ {\rm CL}$  & $95\% \ {\rm CL}$ \\
\hline \hline
~ & \multicolumn{7}{c}{$\Lambda$CDM} & \multicolumn{2}{c}{}\\
\hline
~ & ~ & ~ & ~ & ~ & ~ & ~ & ~ & ~   \\
$\Omega_m$ &0.28 &0.27 & (0.15, 0.4) & (0.11, 0.48) &0.22&0.24 & (0.14, 0.2) & (0.10, 0.35)  \\
~ & ~ & ~ & ~ & ~ & ~ & ~ & ~ & ~  \\
$h$ &  0.74& 0.74& (0.70,0.77) & (0.67, 0.8) &0.75& 0.75& (0.72,0.77) & (0.69,0.8) \\
~ & ~ & ~ & ~ & ~ & ~ & ~ & ~ & ~ \\
\hline\hline\\
~ & \multicolumn{7}{c}{CPL EOS} & \multicolumn{2}{c}{}\\
\hline
~ & ~ & ~ & ~ & ~ & ~ & ~ & ~ & ~ &  \\
$\Omega_m$ & 0.21 &0.18 & (0.12,\, 0.33) & (0.10,\, 0.47) &0.19 &0.20 & (0.12, \,0.27) & (0.11,\,0.44)  \\
~ & ~ & ~ & ~ & ~ & ~ & ~ & ~ & ~  \\
$h$ &  0.74& 0.74 & (0.71,\, 0.77) & (0.66,\, 0.8) &0.74&0.74& (0.70,\,0.77) & (0.67,\,0.8) \\
~ & ~ & ~ & ~ & ~ & ~ & ~ & ~ & ~ \\
$w_{0}$ &-0.65& -0.61 & (-0.77,\,-0.57) & (-0.83,\, -0.52) &-0.64& -0.63 & (-0.75,\, -0.53) & (-0.86,\, -0.51)  \\
~ & ~ & ~ & ~ & ~ & ~ & ~ & ~ & ~ &  \\
$w_{1}$ &-0.1& -0.11 & (-0.3,\,0.15) & (-0.47,\, 0.6) &-0.56& -0.57& (-0.63, \,-0.37) & (-0.69,\, 0.42)  \\
~ & ~ & ~ & ~ & ~ & ~ & ~ & ~ & ~ &  \\
\hline
\end{tabular}}
\end{center}
\caption{Constraints on the cosmological parameters. Columns report the mean $\langle x \rangle$ and median
$\tilde{x}$ values  and the $68\%$ and $95\%$ confidence limits. The analysis is related to
a spatially flat model with the dark energy parametrized through the CPL antsaz.  }
\label{tabcosmofit}
\end{table*}

As far as the cosmological parameters are concerned, it turns out that $\Omega_m^{median}=0.25$,  the range of
confidence  at  ${\rm 1 \sigma}$ is  $\left( 0.13, 0.54\right)$, and
 $\Omega_{\Lambda}^{median}=0.75$,  the range of confidence at  ${\rm 1 \sigma}$ is  $\left( 0.50, 0.87\right)$.
 This result implies that $\Omega_k^{median}=-0.0006$, and that the range of confidence is $\left( -0.00730, 0.00605\right)$, in
 a good agreement with results derived by using the  calibrated  GRBs Hubble diagram.

\begin{table*}
\begin{center}
%\scriptsize
\resizebox{17cm}{!}{
\begin{tabular}{cccccccccc}
\hline
~ & \multicolumn{4}{c}{Fully Bayesian Analysis} & \multicolumn{4}{c}{Local Regression with SNIa} \\
\hline
$Id$ & $\langle x \rangle$ & $\tilde{x}$ & $68\% \ {\rm CL}$  & $95\% \ {\rm CL}$ & $\langle x \rangle$ & $\tilde{x}$ & $68\% \ {\rm CL}$  & $95\% \ {\rm CL}$ \\
\hline \hline
~ & ~ & ~ & ~ & ~ & ~ & ~ & ~ & ~   \\
$a$ & 1.69 &1.71 & (1.64, 1.76) & (1.59, 1.82) &1.74 &1.74 & (1.59, 1.93) & (1.45,2.16)  \\
~ & ~ & ~ & ~ & ~ & ~ & ~ & ~ & ~  \\
$b$ &  52.5& 52.5 & (52.48, 52.55) & (52.44, 52.60) &--& --& -- & -- \\
~ & ~ & ~ & ~ & ~ & ~ & ~ & ~ & ~ \\
$\sigma_{int}$ &0.23& 0.23 & (0.22, 0.25) & (0.20, 0.26) &0.36& 0.37 & (0.33, 0.49) & (0.31, 0.5)  \\
~ & ~ & ~ & ~ & ~ & ~ & ~ & ~ & ~ &  \\
\hline
\end{tabular}}
\end{center}
\caption{Constraints on the calibration parameters. Columns report the mean $\langle x \rangle$ and median
$\tilde{x}$ values  and the $68\%$ and $95\%$ confidence limits. The calibration procedure based on the local
regression technique does not directly provide the zero-point parameter $b$, which can be analytically evaluated as a
function of $(a, \sigma_{int})$ by the Eq. ( \ref{eq:calca}). The fully Bayesian analysis,  furthermore, is related to
a spatially flat model with dark energy parametrized through the CPL antsaz. } \label{tab: fullgrbtab}
\end{table*}

\begin{figure}
\includegraphics[width=8 cm]{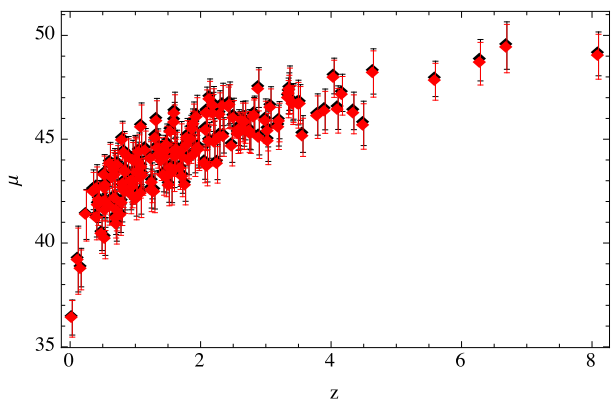}
\caption{Fiducial GRB Hubble diagram, superimposed on the calibrated Hubble diagram.}
\label{hdgrbamati2}
\end{figure}

Also in this case we finally estimate the distance modulus for each $i$\,-\,th GRB in our sample at redshift $z_i$, to
build the fiducial Hubble diagram,  by  using  Eqs. (\ref{eqamati})  and (\ref{lumdist}). It turns out that the fiducial
and calibrated Hubble diagrams are fully statistically consistent, as shown in Fig. \ref{hdgrbamati2}.

\begin{table}
\centering
\resizebox{8 cm}{!}{\begin{minipage}{\textwidth}
        \caption{Derived distance moduli for the low redshift subsample, consisting of the GRBs whose redshifts cross the range spanned by the SNIa.
 Column (1) lists  the GRBs ID. Column (2) lists  the redshift. Column (3) lists the estimated distance modulus $\mu(z)$.
Column (4) lists the error $\sigma_{\mu}$.Column (5) lists $E_{\rm p,i}$.Column (6) lists $\sigma_{E_{\rm p,i}} $.Column (7) 
lists $E_{\rm iso} $.Column (8) lists $\sigma_{E_{\rm iso}} $.}
\label{calibratedHdlow}
 \begin{tabular}{ c|c|c|c||c|c|c|c}
\hline 
 \multicolumn{8}{c}{Calibrated GRBs Hubble diagram: the  low redshift subsample}    \\
\hline
GRBs ID &redshift & $\mu$& $\sigma_{\mu}$ &$E_{\rm p,i} (keV)$ & $\sigma_{E_{\rm p,i}}$ & $E_{\rm iso} (10^{52}\,erg )$ & $\sigma_{E_{\rm iso}}$ \\
\hline \hline\\
060218 & 0.03351 & 35.1195 & 0.933638 & 4.9 & 0.3 & 0.00535399 & 0.000259064 \\
 060614 & 0.125 & 38.6856 & 1.84671 & 55. & 45. & 0.216774 & 0.0867097 \\
 030329 & 0.17 & 38.4667 & 1.03564 & 100. & 23. & 1.47701 & 0.260648 \\
 020903 & 0.25 & 39.9357 & 1.38443 & 3.37 & 1.79 & 0.00244119 & 0.000610299 \\
130427A & 0.3399 & 40.6749 & 0.95616 & 1250. & 150. & 77.0134 & 7.87637 \\
 011121 & 0.36 & 42.9717 & 1.07033 & 1060. & 275. & 7.97042 & 2.18968 \\
 020819B & 0.41 & 40.8059 & 1.1032 & 70. & 21. & 0.693324 & 0.175525 \\
 101213A & 0.414 & 42.8313 & 1.21619 & 440. & 180. & 2.72107 & 0.526658 \\
 990712 & 0.434 & 41.5047 & 0.985045 & 93. & 15. & 0.685192 & 0.131768 \\
 010921 & 0.45 & 41.8452 & 1.0036 & 129. & 26. & 0.966894 & 0.0966894 \\
 091127 & 0.49 & 39.7308 & 0.946699 & 51. & 5. & 1.64622 & 0.176066 \\
 081007 & 0.5295 & 42.6995 & 1.03825 & 61. & 15. & 0.176323 & 0.0176323 \\
 090618 & 0.54 & 39.9036 & 0.925708 & 250.404 & 4.466 & 28.5912 & 0.523898 \\
100621A & 0.542 & 41.4047 & 0.975989 & 146. & 23. & 2.82246 & 0.352807 \\
 060729 & 0.543 & 42.2564 & 1.32376 & 77. & 38. & 0.423384 & 0.0882049 \\
 090424 & 0.544 & 42.0376 & 0.927353 & 249.974 & 3.3196 & 4.06569 & 0.352833 \\
 101219B & 0.55 & 42.5063 & 0.951189 & 108. & 12. & 0.626413 & 0.061759 \\
 050525A & 0.606 & 41.6888 & 0.940851 & 129. & 6.5 & 2.29843 & 0.486207 \\
 110106B & 0.618 & 43.7538 & 1.07668 & 194. & 56. & 0.734033 & 0.0707502 \\
050416A & 0.6528 & 41.8718 & 1.00894 & 22. & 4.5 & 0.10625 & 0.0177083 \\
111209A & 0.677 & 43.7533 & 0.983363 & 520. & 89. & 5.13951 & 0.620285 \\
 080916A & 0.689 & 43.868 & 0.93225 & 208. & 11. & 0.975114 & 0.0886467 \\
 020405 & 0.69 & 42.2854 & 0.928094 & 354. & 10. & 10.6379 & 0.886495 \\
 970228 & 0.695 & 43.1984 & 1.11582 & 195. & 64. & 1.64915 & 0.124129 \\
 991208 & 0.706 & 41.2777 & 0.945272 & 313. & 31. & 22.972 & 1.8626 \\
 111228A & 0.716 & 40.4242 & 0.95484 & 58. & 7. & 2.75042 & 0.26617 \\
 041006 & 0.716 & 41.2868 & 1.01881 & 98. & 20. & 3.10531 & 0.887233 \\
 090328 & 0.736 & 44.393 & 0.9305 & 1157.91 & 55.552 & 14.1806 & 0.997201 \\
 030528 & 0.78 & 40.8528 & 0.974974 & 57. & 9. & 2.22246 & 0.266695 \\
 051022 & 0.8 & 42.3122 & 1.13123 & 754. & 258. & 56.0394 & 5.33709 \\
 100816A & 0.8049 & 44.9536 & 0.94066 & 247. & 20. & 0.711714 & 0.0889643 \\
 110715A & 0.82 & 42.816 & 0.942884 & 220. & 20. & 4.36118 & 0.445018 \\
 970508 & 0.835 & 44.1715 & 1.08963 & 145. & 43. & 0.632199 & 0.133563 \\
 990705 & 0.842 & 42.7009 & 1.09192 & 459. & 139. & 18.7026 & 2.6718 \\
 000210 & 0.846 & 43.8624 & 0.92956 & 753. & 26. & 15.4091 & 1.69233 \\
 040924 & 0.859 & 43.1051 & 1.13207 & 102. & 35. & 0.980131 & 0.0891028 \\
 091003 & 0.8969 & 44.5543 & 0.999868 & 810. & 157. & 10.7036 & 1.78393 \\
 080319B & 0.937 & 42.9071 & 0.931116 & 1261. & 65. & 117.866 & 8.92921 \\
 071010B & 0.947 & 42.1526 & 1.03421 & 88. & 21. & 2.3222 & 0.40192 \\
 970828 & 0.958 & 42.9866 & 1.00157 & 586. & 117. & 30.3759 & 3.57364 \\
 980703 & 0.966 & 44.2503 & 0.957411 & 503. & 64. & 7.41683 & 0.714875 \\
 091018 & 0.971 & 42.7535 & 1.19374 & 55. & 20. & 0.625596 & 0.348546 \\
 980326 & 1. & 43.5588 & 1.33824 & 71. & 36. & 0.500843 & 0.0983799 \\
021211 & 1.01 & 43.7735 & 1.20912 & 127. & 52. & 1.16296 & 0.134188 \\
 991216 & 1.02 & 42.444 & 1.00615 & 648. & 134. & 69.7948 & 7.15844 \\
 080411 & 1.03 & 43.6536 & 0.95959 & 524. & 70. & 16.1999 & 0.984526 \\
 000911 & 1.06 & 44.5421 & 1.00544 & 1856. & 371. & 69.8621 & 14.3307 \\
 091208B & 1.063 & 44.5447 & 0.945957 & 246. & 25. & 2.06018 & 0.179146 \\
 110213B & 1.083 & 43.1534 & 0.974899 & 256. & 40. & 8.3342 & 1.34423 \\
 091024 & 1.092 & 43.1454 & 0.93459 & 396.225 & 25.3132 & 18.3765 & 1.99258 \\
 980613 & 1.096 & 45.4997 & 1.27239 & 194. & 89. & 0.609565 & 0.0986062 \\
 080413B & 1.1 & 44.1225 & 1.08084 & 163. & 47.5 & 1.6137 & 0.268951 \\
 981226 & 1.11 & 43.7092 & 1.27607 & 87. & 40. & 0.807035 & 0.179341 \\
 000418 & 1.12 & 43.2984 & 0.940538 & 284. & 21. & 9.50723 & 1.79382 \\
 061126 & 1.1588 & 45.0298 & 1.09402 & 1337. & 410. & 31.4187 & 3.59071 \\
090926B & 1.24 & 43.924 & 0.945337 & 212. & 21. & 4.13637 & 0.449605 \\
020813 & 1.25 & 42.8407 & 1.05146 & 590. & 151. & 68.3538 & 17.0885 \\
061007 & 1.262 & 43.3478 & 0.963166 & 890. & 124. & 89.9608 & 8.99608 \\
130420A & 1.297 & 42.4236 & 0.945544 & 129. & 13. & 7.74195 & 0.720181 \\
 990506 & 1.3 & 42.8153 & 1.02406 & 677. & 156. & 98.1304 & 9.90306 \\
061121 & 1.314 & 45.6168 & 0.953501 & 1289. & 153. & 23.5036 & 2.70156 \\
071117 & 1.331 & 42.5292 & 1.34174 & 112. & 56. & 5.85526 & 2.70243 \\
 070521 & 1.35 & 44.8194 & 0.949447 & 522. & 55. & 10.8135 & 1.80226 \\
100414A & 1.368 & 44.812 & 0.942611 & 1295. & 120. & 54.987 & 5.40856 \\
120711A & 1.405 & 44.7159 & 0.94466 & 2340. & 230. & 180.405 & 18.0405 \\
100814A & 1.44 & 43.2858 & 0.95932 & 259. & 34. & 15.3438 & 1.80515 \\
 050318 & 1.4436 & 43.8132 & 1.01313 & 115. & 25. & 2.30171 & 0.162473 \\
110213A & 1.46 & 44.1083 & 1.11878 & 224. & 74. & 5.77844 & 0.812593 \\
\hline\hline
\end{tabular}
\end{minipage}
}
\end{table}

\begin{table}
\centering
\resizebox{8 cm}{!}{\begin{minipage}{\textwidth}
        \caption{Derived distance moduli for the high redshift subsample, consisting of the GRBs whose redshifts cross the range spanned by the SNIa.
 Column (1) lists  the GRBs ID. Column (2) lists  the redshift. Column (3) lists the estimated distance modulus $\mu(z)$.
Column (4) lists the error $\sigma_{\mu}$.Column (5) lists $E_{\rm p,i}$.Column (6) lists $\sigma_{E_{\rm p,i}} $.Column (7) 
lists $E_{\rm iso} $.Column (8) lists $\sigma_{E_{\rm iso}} $.}
\label{calibratedHdhigh}
 \begin{tabular}{ c|c|c|c||c|c|c|c}
\hline 
 \multicolumn{8}{c}{Calibrated GRBs Hubble diagram: the high redshift subsample}    \\
\hline
GRBs ID &redshift & $\mu$& $\sigma_{\mu}$ &$E_{\rm p,i} (keV)$ & $\sigma_{E_{\rm p,i}}$ & $E_{\rm iso} (10^{52}\,erg )$ & $\sigma_{E_{\rm iso}}$ \\
\hline \hline\\
 \text{120724A} & 1.48 & 43.9951 & 1.0662 & 69. & 19. & 0.848992 & 0.180637 \\
 10222 & 1.48 & 43.5582 & 0.929143 & 766. & 30. & 84.8992 & 9.03183 \\
 60418 & 1.489 & 45.0134 & 1.043 & 572. & 143. & 13.5498 & 2.70995 \\
 30328 & 1.52 & 42.871 & 0.978916 & 328. & 55. & 38.8622 & 3.61509 \\
 90102 & 1.547 & 45.9243 & 0.946461 & 1174. & 120. & 22.604 & 2.71248 \\
 70125 & 1.547 & 44.0644 & 0.973486 & 934. & 148. & 84.087 & 8.4087 \\
 40912 & 1.563 & 42.781 & 1.69995 & 44. & 33. & 1.35658 & 0.361755 \\
 \text{100728A} & 1.567 & 43.8473 & 0.927328 & 833. & 23. & 86.8265 & 8.13998 \\
 990123 & 1.6 & 44.1755 & 1.0596 & 1724. & 466. & 240.703 & 38.9106 \\
 71003 & 1.604 & 46.5316 & 0.962389 & 2077. & 286. & 38.2795 & 4.52476 \\
 \text{090418A} & 1.608 & 46.8731 & 1.03718 & 1567. & 384. & 17.1951 & 2.71502 \\
 \text{110503A} & 1.613 & 44.6927 & 0.948442 & 551. & 60. & 20.8167 & 1.81015 \\
 990510 & 1.619 & 44.3532 & 0.946002 & 423. & 42. & 18.1031 & 2.71546 \\
 80605 & 1.64 & 45.0375 & 0.958753 & 766. & 55. & 28.0685 & 14.4869 \\
 80928 & 1.6919 & 43.2837 & 1.03666 & 95. & 23. & 3.98686 & 0.906105 \\
 91020 & 1.71 & 45.675 & 0.960875 & 507.228 & 68.208 & 8.4289 & 1.0876 \\
 \text{100906A} & 1.727 & 43.2599 & 0.973643 & 289. & 46. & 29.916 & 2.71963 \\
 \text{120119A} & 1.728 & 44.3889 & 0.946023 & 496. & 50. & 27.1967 & 3.62623 \\
 \text{110422A} & 1.77 & 42.9739 & 0.944986 & 421. & 42. & 79.8218 & 8.1636 \\
 \text{120326A} & 1.798 & 44.5549 & 0.94189 & 152. & 14. & 3.26663 & 0.272219 \\
 \text{080514B} & 1.8 & 45.3824 & 0.948821 & 627. & 65. & 18.1484 & 3.62968 \\
 \text{090902B} & 1.822 & 44.7661 & 0.925467 & 2187. & 31. & 292.271 & 9.07675 \\
 \text{110801A} & 1.858 & 45.1695 & 1.23345 & 400. & 171. & 10.897 & 2.72425 \\
 60908 & 1.8836 & 46.2739 & 1.28842 & 553. & 260. & 7.1761 & 1.90757 \\
 20127 & 1.9 & 45.7854 & 1.13272 & 290. & 100. & 3.72504 & 0.109025 \\
 60814 & 1.9229 & 44.6651 & 1.11463 & 751. & 246. & 56.7086 & 5.27099 \\
 803190 & 1.95 & 46.5089 & 1.08858 & 906. & 272. & 14.9089 & 2.99996 \\
 81008 & 1.9685 & 44.6084 & 0.999689 & 261. & 52. & 10.002 & 0.909269 \\
 30226 & 1.98 & 44.5553 & 1.02204 & 289. & 66. & 12.7314 & 1.36408 \\
 \text{081203A} & 2.05 & 46.8257 & 1.31834 & 1541. & 756. & 31.853 & 11.8311 \\
 926 & 2.07 & 43.9293 & 0.936188 & 310. & 20. & 28.5827 & 6.18989 \\
 80207 & 2.0858 & 44.6893 & 1.56663 & 333. & 222. & 16.3877 & 1.82086 \\
 \text{061222A} & 2.088 & 45.8633 & 0.983381 & 874. & 150. & 30.0449 & 6.37315 \\
 \text{100728B} & 2.106 & 46.3176 & 0.965919 & 323. & 47. & 3.55142 & 0.364248 \\
 90926 & 2.1062 & 43.7433 & 0.925125 & 900.798 & 7.02001 & 228.014 & 2.27338 \\
 11211 & 2.14 & 44.7932 & 0.957566 & 186. & 24. & 5.73888 & 0.637653 \\
 71020 & 2.145 & 47.4073 & 0.984835 & 1013. & 160. & 10.0208 & 4.5549 \\
 509220 & 2.199 & 46.4216 & 1.06071 & 415. & 111. & 5.55991 & 1.82292 \\
 \text{121128A} & 2.2 & 44.9272 & 0.931098 & 243. & 13. & 8.65897 & 0.820323 \\
 80804 & 2.2045 & 46.8579 & 0.931104 & 810. & 45. & 12.0319 & 0.546906 \\
 \text{110205A} & 2.22 & 45.2389 & 1.20041 & 757. & 305. & 48.3171 & 6.38151 \\
 81221 & 2.26 & 43.8782 & 0.929903 & 284. & 14. & 31.9195 & 1.82397 \\
 \text{130505A} & 2.27 & 45.029 & 0.935913 & 2030. & 150. & 346.586 & 27.3621 \\
 60124 & 2.296 & 45.502 & 1.15448 & 784. & 285. & 43.7896 & 6.38599 \\
 21004 & 2.3 & 46.2112 & 1.2461 & 266. & 117. & 3.46681 & 0.456159 \\
 \text{051109A} & 2.346 & 46.8641 & 1.16259 & 539. & 200. & 6.84516 & 0.730151 \\
 \text{080413A} & 2.433 & 46.867 & 1.09682 & 584. & 180. & 8.58559 & 2.10073 \\
 90812 & 2.452 & 47.364 & 1.13965 & 2000. & 700. & 47.5021 & 8.22152 \\
 \text{120716A} & 2.486 & 44.8287 & 0.944518 & 397. & 40. & 30.1538 & 0.274125 \\
 \text{130518A} & 2.488 & 45.181 & 0.942494 & 1382. & 130. & 192.805 & 18.2753 \\
 81121 & 2.512 & 45.5879 & 0.934876 & 608. & 42. & 32.3534 & 3.65576 \\
 81118 & 2.58 & 44.4939 & 0.943808 & 203.308 & 20. & 13.9705 & 0.888677 \\
 80721 & 2.591 & 46.1254 & 0.958853 & 1741. & 227. & 133.515 & 22.8622 \\
 50820 & 2.615 & 45.9102 & 1.00661 & 1325. & 277. & 103.356 & 8.23186 \\
 30429 & 2.65 & 45.6525 & 1.00258 & 128. & 26. & 2.28721 & 0.274465 \\
 \text{120811C} & 2.671 & 45.3823 & 0.943181 & 198. & 19. & 6.40514 & 0.640514 \\
 \text{080603B} & 2.69 & 46.1013 & 1.16024 & 277. & 100. & 6.03994 & 3.05658 \\
 91029 & 2.752 & 45.5087 & 1.07352 & 230. & 66. & 7.96513 & 0.823979 \\
 81222 & 2.77 & 46.0964 & 0.9302 & 630.344 & 31.2156 & 27.3822 & 2.74693 \\
 50603 & 2.821 & 46.6406 & 0.937652 & 1333. & 107. & 64.1164 & 4.57975 \\
 \text{110731A} & 2.83 & 46.6737 & 0.930181 & 1164. & 58. & 49.4641 & 4.58001 \\
 \text{111107A} & 2.893 & 47.598 & 1.08191 & 420. & 124. & 3.7571 & 0.54982 \\
 50401 & 2.8983 & 45.3038 & 1.02857 & 467. & 110. & 37.5723 & 7.33118 \\
 \text{090715B} & 3. & 46.1498 & 1.10789 & 536. & 172. & 23.8409 & 3.66783 \\
 80607 & 3.036 & 46.0503 & 0.959184 & 1691. & 226. & 199.938 & 11.0058 \\
 81028 & 3.038 & 44.8963 & 1.19331 & 234. & 93. & 18.3432 & 1.83432 \\
 \text{060607A} & 3.075 & 46.7497 & 1.03846 & 478. & 118. & 11.9256 & 2.75205 \\
 20124 & 3.2 & 45.7872 & 1.11734 & 448. & 148. & 28.4571 & 2.75391 \\
 60526 & 3.22 & 45.5887 & 1.00027 & 105. & 21. & 2.7542 & 0.367227 \\
 80810 & 3.35 & 47.5973 & 0.954116 & 1470. & 180. & 47.7705 & 5.51198 \\
 \text{110818A} & 3.36 & 47.7168 & 1.01118 & 1116. & 240. & 26.6425 & 2.75612 \\
 30323 & 3.37 & 47.4274 & 1.22111 & 270. & 113. & 2.94001 & 0.918752 \\
 971214 & 3.42 & 47.0429 & 0.996013 & 685. & 133. & 22.0553 & 2.75691 \\
 60707 & 3.424 & 47.0792 & 1.05247 & 274. & 72. & 4.31924 & 1.10278 \\
 60306 & 3.5 & 46.7831 & 1.23155 & 315. & 135. & 7.63024 & 1.01124 \\
 980329 & 3.5 & 45.2884 & 0.974999 & 1096. & 176. & 266.599 & 53.3198 \\
 60115 & 3.5328 & 46.978 & 1.1415 & 297. & 102. & 5.88442 & 3.7697 \\
 90323 & 3.57 & 45.8457 & 0.986639 & 1901. & 343. & 437.725 & 53.3362 \\
 \text{130514A} & 3.6 & 45.6286 & 1.09246 & 497. & 152. & 52.4234 & 9.19708 \\
 \text{130408A} & 3.758 & 47.5053 & 0.963238 & 1000. & 140. & 34.9716 & 6.44214 \\
 \text{120802A} & 3.796 & 46.1613 & 1.12782 & 274. & 93. & 12.8862 & 2.76133 \\
 60210 & 3.91 & 46.6447 & 1.11229 & 574. & 187. & 32.2294 & 1.84168 \\
 \text{120909A} & 3.93 & 47.0881 & 1.17098 & 1276. & 483. & 87.4863 & 10.13 \\
 60206 & 4.0559 & 48.3284 & 1.27013 & 410. & 187. & 4.14595 & 1.93478 \\
 90516 & 4.109 & 46.899 & 1.19876 & 971. & 390. & 71.8762 & 13.8223 \\
 \text{120712A} & 4.1745 & 47.4784 & 1.00188 & 641. & 130. & 21.1989 & 1.84338 \\
 \text{080916C} & 4.35 & 47.145 & 0.960048 & 2760. & 369. & 406.692 & 85.765 \\
 131 & 4.5 & 46.1465 & 1.22279 & 987. & 416. & 183.6 & 32.2915 \\
 90205 & 4.6497 & 49.1939 & 1.1254 & 214. & 72. & 0.8307 & 0.2769 \\
 60927 & 5.46 & 47.1793 & 1.06039 & 275. & 75. & 12.0218 & 2.77427 \\
 50904 & 6.295 & 49.5673 & 1.13213 & 3178. & 1094. & 133.364 & 13.8921 \\
 80913 & 6.695 & 49.788 & 1.31553 & 710. & 350. & 9.1742 & 2.68739 \\
 90423 & 8.1 & 49.2684 & 1.0345 & 410. & 100. & 8.81815 & 2.0653 \\
 \text{090429B} & 9.3 & 50.0158 & 0.959439 & 433. & 58. & 6.69027 & 1.30089 \\
 \hline \hline
\end{tabular}
\end{minipage}}
%\end{center}
\end{table}

\section{Discussion and conclusions}

The  $E_{\rm p,i}$ -- $E_{\rm iso}$ correlation is one of the most intriguing properties of the long GRBs, with
significant implications  for the use of GRBs as cosmological probes. Here we
explored the Amati relation in a way independent of the cosmological model. Using the recently
updated data set of 162 high-redshift GRBs, we  applied a local regression technique to estimate the distance modulus
using the recent Union SNIa sample (Union2.1). The derived calibration parameters are statistically fully consistent
with the  results of our  previous work  \citep[][]{MEC11}.  Moreover,  we tested the validity of the commonly adopted working
hypothesis that the GRB Hubble diagram is slightly affected by  redshift dependence of the $E_{\rm p,i}$ -- $E_{\rm iso}$ correlation.
As a first  qualitative and simplified approach we considered a {\it lower redshift sample}  of $97$ bursts with $z\leq 2$, and a  
higher redshift sample of 67 burst with $z>2$.  We estimated the  cosmological parameters  for  flat $\Lambda CDM$  and  CPL dark 
energy EOS cosmological models, considering
the two subsamples separately, and compared the results. Even when the bursts belonging to these two samples had different  environment 
conditions, we found no significant indications that a spurious z-evolution of the slope affected the  cosmological fit. 
Moreover, to quantify this redshift dependence, we used two different approaches. First, we evaluated the Spearman rank 
correlation coefficient, $C(z,y)$ by applying  a jacknife re--sampling method by evaluating $C(z, y)$ for $N-1$ samples obtained 
by excluding one GRB at a time, and we adopted the mean value and the standard deviation  to estimate $C(z,y)$.  
$ C(z,E_{p,i})= 0.299 \pm 0.004$, and $ C(z, E_{iso})= 0.278 \pm 0.004$, which indicates a negligible evolution of the correlation.
Moreover, we  also implemented an alternative method, assuming that  the redshift evolution can be parametrized by simple power-law 
functions: $g_{iso}(z)=\left(1+z\right)^{k_{iso}}$ and $g_{p}(z)=\left(1+z\right)^{k_{p}}$ and that the correlation holds for the 
de-evolved quantities  $E_{\rm iso}^{'} =\displaystyle\frac{E_{\rm iso}}{g_{iso}(z)}$ and $E_{\rm p,i}^{'} =\displaystyle\frac{E_{\rm p,i}}{g_{p}(z)}$ . 
In this case,  we rewrote an effective 3D $E_{\rm iso}-E_{\rm p,i}$ correlation, with included the evolutionary terms. Since we were 
interested in the implications of possible evolutionary  effects of the GRB Hubble diagram and to simplify the fit, we introduced an
auxiliary variable  $\alpha= a k_p$. A null value for $k_{iso}$ and $\alpha$ is strong evidence for a lack of any evolution.
To fit the coefficient of our 3D correlation we constructed a 3D Reichart likelihood and used a  MCMC method running five parallel chains and
using the Gelman-Rubin
convergence test. The $E_{\rm iso}$ and  $E_{\rm p,i}$ correlation shows at this stage only weak indications of evolution.
The derived calibration parameters were used to construct an updated  calibrated GRB Hubble diagram, which we adopted as
a tool  to constrain the cosmological models and to investigate the dark energy EOS. In particular, we searched for
any indications that $w(z)\neq-1$, which reflects  the possibility of a deviation  from the $\Lambda$CDM cosmological
model. To accomplish this task, we focused on the CPL parametrization as an example.  To efficiently sample the
cosmological parameter space, we again used a  MCMC method. At $1 \sigma$ level we found  indications for a time
evolution of the EOS  in the considered parametrization; we
conclude that the  $\Lambda$CDM model is not favored even though it is not ruled out by these observations so far.
Moreover, for $w=-1$ we also performed our analysis assuming a non-zero spatial curvature, adding a Gaussian
prior on $\Omega_k$ centered on the value provided by the Planck data \citep[][]{PlanckXXVI}. The GRB Hubble diagram alone
provides $\Omega_k^{median}= -0.00046$  for  the range of confidence at  ${\rm 1 \sigma}$  $\left( -0.007, 0.0064
\right)$.  Finally,  to investigate the reliability of the $E_{\rm
p,i}$ -- $E_{\rm iso}$ correlation in greater detail, we also
used a different method to simultaneously extract the correlation
coefficients and the cosmological parameters of the model from the observed quantities.  To illustrate this method we
assumed here, by way of an example, only  the particular case of a non-flat $\Lambda$CDM model, as already done in the
previous analysis. This analysis simultaneously determines the cosmological parameters and the correlation
coefficients. Although the calibration procedure is different (since we now fit for the cosmological parameters
as well), it is nevertheless worth comparing this determination of $(a, b, \sigma_{int})$ with the one  obtained in
the previous analysis based on the SNIa sample. It is evident that although the median values change, the  $95\%$
confidence levels are in full agreement so that we cannot find any statistically significant difference. This means
that the high-redshift GRBs can be used as cosmological probes, mainly in a redshift region, $z\geqslant 3$, which is 
unexplored by SNIa and BAO samples, and that both the calibration technique based on a  local regression with SNIa and a
fully Bayesian approach are reliable.  As a final remark we note that our results for the cosmological parameters are statistically 
consistent with those previously obtained by other teams, as shown in Table (\ref{cosmo_oth}), where we display some of the most 
recent measurements  of cosmological parameters obtained with GRBs, even if following a slightly different approach. The main
peculiarity in our analysis consists of the procedure used to build up the dataset, consisting of 162 objects,  as discussed in Sect.
(\ref{GRBdata}) (see also \citep[][]{DishaLz14, Sawant16}), even more than the specific statistical analysis, which follows a
Bayesian approach through the implementation of the MCMC. Because the reliability of GRBs as distance indicators has not yet been clearly proved and because
it constitutes an independent  topic in the field of GRB research, it is not meaningless to investigate the reliability of
the $E_{\rm p,i}$ -- $E_{\rm iso}$ correlation whenever new and improved datasets are available.  In the near future we intend 
to enhance the cosmological analysis by using the high-redshift GRB Hubble diagram to test different cosmological models, where 
several  parameterizations of the dark energy EOS are
used,  but a cosmographic approach will also be implemented,
which will update the analysis performed in \citep[][]{MECP12} to check whether with this new updated dataset the estimates of 
the jerk and the dark energy  parameters will confirm deviations from $\Lambda$CDM cosmological model, as has been indicated in our previous analysis.
\begin{table}
\begin{center}
\resizebox{9 cm}{!}{
\begin{tabular}{c|c|c|c}
\hline
Data & Cosmological model & Constraint & Reference \\
\hline

GRB & flat $\Lambda$CDM &$\Omega_m=0.23_{-0.04}^{+0.06} $&\cite{WangW} \\
GRB & flat $\Lambda$CDM &$\Omega_m=0.302\pm0.142$ &\cite{Lin16b}\\
GRB+SNeIa+Bao+Planck & flat $\Lambda$CDM&$\Omega_m=0.2903^{+0.0109}_{-0.0106}$& \cite{Liu15}\\
 GRB & flat $\Lambda$CDM &$ \Omega_m= 0.29^{+0.23}_{-0.15}$ & \cite{Izzo15}\\
\hline
\end{tabular}}
\caption{ Some of the most recent  constraints of cosmological parameters by GRBs.}
\label{cosmo_oth}
\end{center}
\end{table}

\subsection*{Acknowledgments}
MD is grateful to the INFN for financial support through the Fondi FAI GrIV.
EP acknowledges the support of INFN Sez. di Napoli  (Iniziative Specifica QGSKY and TEONGRAV).
LA  acknowledges suppo\ rt by the Italian Ministry for Education,
University and Research through PRIN MIUR 2009 project on "Gamma ray
bursts: from progenitors to the physics of the prompt emission process." (Prot. 2009 ERC3HT).

\begin{onecolumn}

\bibliographystyle{apa}

\end{onecolumn}

\end{document}